\documentclass{aa} 

\usepackage{graphicx}
\usepackage{txfonts}
\usepackage{hyperref}
\usepackage{bm}

\newcommand\id{{\mathrm d}}

\newcommand\br{{\bm r}}

\newcommand\bu{{\bm u}}

\newcommand\Cref{C_{\text{ref}}}
\newcommand\cref{\Cref}

\bibpunct{(}{)}{;}{a}{}{,} 

\begin{document} 

\title{Direct driving of simulated planetary jets by upscale energy transfer}

\titlerunning{Direct driving of simulated planetary jets}

\author{
        Vincent G. A. Böning\inst{1}
        \and
        Paula Wulff\inst{1,2}
        \and
        Wieland Dietrich\inst{1}
        \and
        Johannes Wicht\inst{1}
        \and
        Ulrich R. Christensen\inst{1}
}

\institute{
        {Max-Planck-Institut f\"ur Sonnensystemforschung, Justus-von-Liebig-Weg 3, 37077 G\"ottingen, Germany}\\
        \email{boening@mps.mpg.de}
        \and
        {Georg-August-Universit\"at G\"ottingen, Friedrich-Hund-Platz 1, 37077 G\"ottingen, Germany}
        }

\authorrunning{Böning et. al.}

\date{Received ???; Accepted ???}

\abstract
{The precise mechanism that forms jets and large-scale vortices on {the} giant planets is unknown. An inverse cascade has been suggested by several studies. Alternatively, energy may be directly injected by small-scale convection.}
{{Our} aim is to clarify whether an inverse cascade {feeds zonal jets and large-scale eddies} in a system of rapidly rotating, {deep, geostrophic} spherical-shell convection.}
{We analyze the nonlinear scale-to-scale transfer of kinetic energy in such simulations as a function of the azimuthal wave number, $m$.}
{We find that the main driving of the jets is associated with upscale transfer directly from the small convective scales to the jets. This transfer is very nonlocal in spectral space, bypassing large-scale structures. The jet formation is thus not driven by an inverse cascade. Instead, it is due to a direct driving by Reynolds stresses, statistical correlations of velocity components of the small-scale convective flows. Initial correlations are caused by the effect of uniform background rotation and shell geometry on the flows and provide a seed for the jets. While the jet growth suppresses convection, it increases the correlation of the convective flows, which further amplifies the jet growth until it is balanced by viscous dissipation. To a much smaller extent, energy is transferred upscale to large-scale vortices directly from the convective scales, mostly outside the tangent cylinder. There, large-scale vortices are not driven by an inverse cascade either. Inside the tangent cylinder, the transfer to large-scale vortices is even weaker, but more local in spectral space, leaving open the possibility of an inverse cascade as a driver of large-scale vortices. In addition, large-scale vortices receive kinetic energy from the jets via forward transfer. 
We therefore suggest a jet instability as an alternative formation mechanism of large-scale vortices. 
{Finally, we find that the jet kinetic energy scales approximatively as $\ell^{-5}$, the same as for the so-called zonostrophic regime.}
}
{}


\maketitle

\section{Introduction}
\label{secIntro}

The flow at the surfaces of the giant planets Jupiter, Saturn,
Uranus, and Neptune is dominated by strong zonal jets. 
In addition, Jupiter's atmosphere features prominent eddies over a larger range of sizes, the great red spot being the largest. Recent observations of the polar regions by the Juno mission also revealed that Jupiter's poles are surrounded by larger cyclonic eddies \citep[][]{Adriani2018,Gavriel2021}. Saturn's polar hexagon may also be explained by eddies shaping a zonal wind jet \citep[][]{Yadav2020}. 

Until the Juno mission, it was an open question as to whether the jets are features of a thin weather layer, as on Earth, or if they reach deep into the planet. 
The recent gravity measurements of the Juno and Cassini missions show that the winds reach depths of about 3000 km on Jupiter \citep[][]{Guillot2018,Iess2018,Kaspi2018} and about 8000 km on Saturn \citep[][]{Galanti2019,Iess2019}. The jets are therefore deep and not restricted to a weather layer, but the questions remain as to which mechanism prevents them from penetrating even deeper and at precisely what depth they decay \citep[e.g.,][and references therein]{Kong2018,Wicht2020,Dietrich2021}. Recently, it has been suggested that a stably stratified layer could be responsible \citep[][]{Christensen2020,Wicht2020b}.

Although deeply penetrating, the jets could still be driven by {motions} in a shallow weather layer \citep[e.g.,][]{Showman2007,Schneider2009,Lian2010,Read2020,Cabanes2020}. Indeed, the dynamics of rotating systems predicts that the jets organize on geostrophic cylinders with minimal variation along the rotation axis, independent of where along the cylinders the flow is driven.

\begin{figure}
    \centering
    \includegraphics[width=\linewidth]{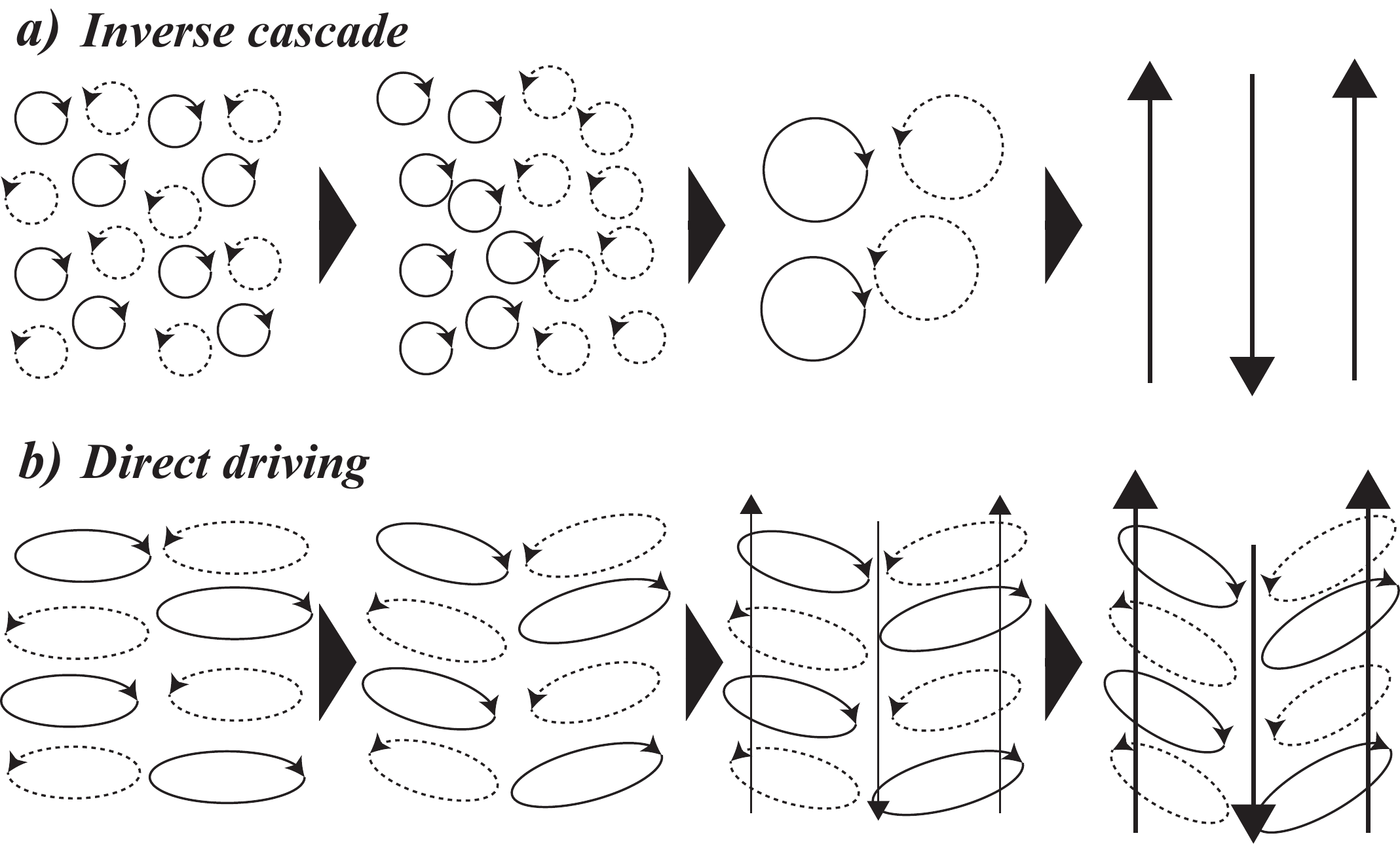}
    \caption{Schematic illustration of two possible mechanisms for jet driving. The top row  (a) shows the {inverse} cascade idea, where eddies merge until the jet scale is reached. The first step in this process illustrates the required anisotropy necessary to prefer either cyclonic or anticyclonic eddies, here illustrated by a sorting of both eddy types. The bottom row (b) shows the direct driving via Reynolds stresses, where an initial tilt of convective features (second column) yields weak zonal flows that then amplify the tilt until a viscous balance is reached.
}
    \label{figIllustration}
\end{figure}

{Traditionally, there were} two competing theories regarding how giant planetary jets are driven. Both {attempt to} explain how kinetic energy is transported from a small convective driving scale to the jet scale. The first theory invokes a kinetic energy cascade and the second a direct driving of jets by small-scale convection (see Fig.~\ref{figIllustration}). {Recently, some evidence has accumulated in favor of the direct driving idea \citep[][]{Galperin2019ZonalJetsBook}, but there is still no final conclusion on the precise driving mechanism of the jets.}

Cascades are known to transport energy from the driving scale to the scales where the energy is dissipated. In a forward cascade, the energy is transferred downscale to smaller scales \citep[][]{Kolmogorov1941}, which is more common for three-dimensional dynamics \citep[see, e.g.,][for a review on cascades]{Alexakis2018}. An inverse cascade transfers energy upscale to larger scales and is more common for two-dimensional dynamics \citep{Kraichnan1967}. Responsible for any energy exchange between different scales is the nonlinear inertial (or advective) term in the Navier-Stokes equation. The range of scales over which this transport to the dissipative scale dominates is therefore called the inertial range. A cascade has two characteristic properties \citep[e.g.,][p. 104]{Frisch1995TurbulenceBook}. First, the dynamics should be scale-invariant in the inertial range. This includes a constant scale-to-scale flux of energy. Second, the word ``cascade'' implies a continuous transfer between scales, which should be local in wavenumber space, meaning that several intermediate steps occur between the driving and dissipation scales and that eddy-eddy interactions play a major role \citet[][p. 66]{Richardson1922}.

An inverse cascade could therefore explain the transport of energy from the driving scale to the large jet scale (see Fig.~\ref{figIllustration}a). This relates to the idea of eddy merging: starting at the driving scale, eddies would continue to merge to form larger and larger features. One idea is that the merging would stop at the so-called Rhines length scale, where turbulent inertial dynamics {give} way to Rossby waves \citep[][]{Rhines1975}. {In a somewhat similar scenario called the zonostrophic regime \citep[e.g.,][]{Galperin2006,Sukoriansky2007}, a two-dimensional inverse cascade operates at intermediate length scales. These studies assume a beta plane approximation or forced turbulence on a rotating spherical surface. Since this inverse cascade is isotropic in spectral space, it does not favor the creation of zonal structures. At scales larger than the Pelinovsky-Vallis-Maltrud scale \citep[][see also \citealp{Sukoriansky2007}]{Pelinovsky1978,Vallis1993}, the inverse cascade becomes anisotropic; due to the presence of the beta effect, kinetic energy is preferentially transferred to zonal wave numbers, which leads to the driving of jets. Again, the inverse cascade stops at the Rhines scale in this scenario.} 
Numerical experiments have shown that the anisotropy caused by the variation in Coriolis force with latitude {results in a spectral anisotropization of the inverse cascade, which} helps explain why flows at the Rhines scale preferably organize in axisymmetric jets {rather than in large eddies \citep[see][for overviews]{Vasavada2005,Sukoriansky2007}.} 
{The} Rhines scale indeed successfully describes the jet width for Jupiter, for Saturn, for laboratory experiments, and for {deep-shell} numerical 
simulations \citep[e.g.,][]{Heimpel2007,Gastine2014}. {This} inverse cascade picture of jet formation is traditionally rather connected to two-dimensional or shallow general circulation models \citep[e.g.,][]{Rhines1975,Williams1978,Vallis1993,Showman2007,Cabanes2020}. {\citet[][]{Young2017} measured the upscale transfer of kinetic energy in observations of flows in Jupiter's atmosphere obtained from cloud tracking by the Cassini spacecraft and find considerable evidence for the eddy-eddy transfer being due to an inverse cascade.}

The eddies on Jupiter seem to be shallower, yet also penetrate deeper than the weather layer \citep[e.g.,][]{Parisi2021}. Eddy merging and thus a cascade provide a 
possible driving scenario for these features as well. An inverse cascade to large-scale vortices could be expected from three-dimensional simulations of rapidly rotating Rayleigh-Bénard convection \citep[e.g.,][]{Rubio2014,Favier2014,Kunnen2016,Guervilly2017,Novi2019,Maffei2021}, where the large-scale flows have a strong quasi-two-dimensional component, similar to the spherical-shell case. These authors, however, employ a slightly less strict definition of a cascade by not including the necessity of local interactions. In agreement with these simulations, \citet[][]{Siegelman2022} find an upscale transfer to large scales in the Juno data of Jupiter's polar vortices. The locality of the transfer has, however, not been constrained, and it is an open question as to whether Jupiter's polar vortices are driven by an inverse cascade.

An alternative explanation to the cascade model is a direct driving of the jets by Reynolds stresses (see Fig.~\ref{figIllustration}b). Reynolds stresses describe a statistical correlation of small-scale flow components over the jet scale. This picture is usually connected to deep-shell simulations of rotating convection \citep[e.g.,][]{Aurnou2001,Christensen2001,Heimpel2005,Kaspi2009,Gastine2012,Yadav2020}. {It is consistent with cloud tracking observations of the surface flows of Jupiter \citep[e.g.,][]{Salyk2006,Ingersoll2021,Duer2021}, which show that the observed Reynolds stresses are consistent with a direct flux of angular momentum from the eddies into the jets. The corresponding upscale energy flux from the eddies to the jets is about three times stronger than the eddy-eddy upscale flux \citep[][]{Young2017} and thus supports the idea of primarily a direct driving. To achieve a correlation that drives the jets, azimuthal and meridional flows should be sufficiently correlated. The latitudinal scale of this correlation should reflect the jet width.} A straightforward way to achieve this is a tilt of geostrophic convective features. At mild parameters, these features may still assume the form of convective columns, while
at more extreme parameters, the columns may only exist in
a statistical sense. 
The shear provided by the jets is a very efficient way to cause the tilt. A small initial flow can thus lead to runaway growth until viscous effects provide a balance. These ideas were introduced by Busse under the name mean flow instability \citep[see, e.g.,][for an overview]{Busse2002}. Eddy-mean flow interactions play the main role in this picture, which is supported by quasi-linear simulations  \citep[e.g.,][]{Tobias2011,Srinivasan2012,Marston2016}.

What has been missing is an analysis of the scale-to-scale transfer of kinetic energy, which would allow us to determine whether an inverse cascade is operating in deep-shell simulations of rapidly rotating convection. We provide such an analysis here. Our main aim is therefore to answer the question of whether in these simulations energy is transferred into the jets in steps involving intermediate- and large-scale eddies or whether the transfer occurs directly from small-scale convection. Such an analysis would indicate how a deep driving of the jets could be constrained observationally. {Since buoyancy does not drive jets, an upscale transfer of kinetic energy is in any case expected from the presence of jets and has been measured in simulations of convection in a spherical shell \citep[][]{Reshetnyak2013}, leaving both a direct driving and a driving by an inverse cascade as possibilities.}

We restrict ourselves to studying the transfer between different azimuthal wave numbers, {$m$. Our analysis addresses the question of how kinetic energy, which is driven by buoyancy at typical convective wave numbers $m>0$, finds its way into the axisymmetric jets, which have wave {number} $m=0$. In particular, we are interested in the question of which wave numbers drive the jets and whether there is a cascade involving more than one step from the convective driving scale to $m=0$. We leave a detailed discussion of the jet width and the Rhines scale \citep[][]{Rhines1975,Gastine2014} to future studies, for which an analysis of energy transfer between harmonic degrees, $\ell$, is more appropriate \citep[e.g.,][]{Young2017}}.

\section{Numerical simulations of deep spherical-shell convection}

\label{secMethods}

\subsection{Governing equations and numerical implementation}

We analyzed the results of direct numerical simulations of convection-driven flow in a rotating spherical shell with inner radius $r_i$ and outer radius $r_o$ using the pseudo-spectral magneto-hydrodynamic code MagIC \citep[][version 5.10]{Wicht2002,Gastine2012,Lago2021}. We solved the hydrodynamic equations in the anelastic approximation in a nondimensional form. The momentum equation in a corotating frame is
\begin{align}
\dfrac{\partial \vec{u}}{\partial t}+\vec{u}\cdot\vec{\nabla}\vec{u}
= -\vec{\nabla}\left({\dfrac{p}{\tilde{\rho}}}\right) - \dfrac{2}{E}\vec{e_z}\times\vec{u}
+ \dfrac{Ra}{Pr}\tilde{g} \,S\,\vec{e_r}
+ \dfrac{1}{\tilde{\rho}} \vec{\nabla}\cdot \mathcal{S}, \label{eqNSE}
\end{align}
the continuity equation is
\begin{align}
\vec{\nabla}\cdot\tilde{\rho}\vec{u}=0,    \label{eqMassCons}
\end{align}
and the entropy equation is
\begin{align}
\tilde{\rho}\tilde{\Theta}\left(\dfrac{\partial S}{\partial t} +
\vec{u}\cdot\vec{\nabla} S\right) =&
\dfrac{1}{Pr}\vec{\nabla}\cdot\left(\tilde{\rho}\tilde{\Theta}\vec{\nabla} S\right) + \dfrac{Pr\,Di}{Ra}\Phi_\nu. \label{eqs}
\end{align}
 The nondimensional rate-of-strain tensor $\mathcal{S}$ in Eq.~\eqref{eqNSE} is given by
 \begin{align}
\mathcal{S}_{ij}  &= 2\tilde\rho e_{ij} - \dfrac{2\tilde\rho}{3}\,(\vec{\nabla}\cdot\vec{u}) \, \delta_{ij}, \label{eqSij}\\
e_{ij} &=\dfrac{1}{2}\left(\dfrac{\partial u_i}{\partial x_j}+\dfrac{\partial
        u_j}{\partial x_i}\right), \label{eqeij}
\end{align}
and the viscous dissipation into heat is
\begin{align}
\Phi_\nu &= 2\tilde\rho\left[e_{ij}e_{ji}-\dfrac{1}{3}\left(\vec{\nabla}\cdot\vec{u}\right)^2\right] \label{eqPhinu}.
\end{align}
Here $\vec{u}$ is the velocity field, $p$ the modified pressure (including the centrifugal force) and $S$ the entropy. Further,  $\vec{e_r}$ and $\vec{e_z}$ are the unit vector in the radial direction and along the axis of rotation, respectively. The quantities $\tilde g$, $\tilde \rho$ and $\tilde \Theta$ are purely radial profiles of gravity, density and temperature characterizing the hydrostatic and adiabatic  background state on which the fluctuations evolve according to Eqs.~\eqref{eqNSE}-\eqref{eqs}. 
We use a spherical shell with a radius ratio of $r_i/r_o=0.7$ and assume a polytropic ideal gas with a polytropic index of $n=2$. The contribution of the shell mass to the total mass can be neglected and thus the gravity profile is $\tilde{g}= r_o^2/r^2$. In most of this work, we use spherical polar coordinates $(r,\theta,\varphi)$, while at some locations, we also employ cylindrical coordinates ($s,z,\varphi$). 
Equations~\eqref{eqNSE}-\eqref{eqs} were non-dimensionalized by using the shell thickness $d=r_o-r_i$ as a length scale and the viscous timescale $\tau_{\rm visc}=d^2/\nu$ as a timescale. The nondimensional velocity $\bu$ is therefore expressed in terms of a Reynolds number.

We fix the nondimensional input parameters, such as Ekman ($E$), Rayleigh ($Ra$), Prandtl ($Pr$), and dissipation ($Di$) numbers to
\begin{align}
E &= \frac{\nu}{\Omega d^2} = 3\times10^{-5}, \\
Ra &= \frac{\alpha_o g_o \Theta_o d^4 }{c_p \kappa \nu} \, \Big\vert\frac{d S}{d r}\Big\vert_o = 3\times 10^8, \\
Pr &= \frac{\nu}{\kappa} = 0.5, \\
Di &= \dfrac{\alpha_o d \tilde{g}_o}{c_p}  = 1\ , \label{eqDi2}
\end{align}
where $\Omega$ is the rotation rate, $\vert d S / d r\vert_o$ is the entropy gradient at the outer boundary (the same as at the inner boundary), $\nu$ is the viscosity, and $\kappa$ is the thermal diffusivity. In addition, $\alpha_o$, $\Theta_o$, and $g_o$ are the reference values of thermal expansion coefficient, temperature, and gravity at the outer boundary. For this choice of parameters, the total density contrast across the shell is $\rho_i/\rho_o \approx 6$ and the convection is quite vigorous, yet still rotationally constrained ($Ra E^2/Pr < 1$). Similarly for both boundaries, we apply free-slip conditions for the flow and fixed entropy flux conditions, which are relatively standard for simulating planetary jet formation \citep[e.g.,][]{Heimpel2016}.

\subsection{Rapidly rotating convection {with moderately turbulent geostrophic flows}}
\label{secGeostrophic}

We analyze{d} a simulation that has been integrated long enough to reach a statistically steady state. In particular, we made sure that the zonal flows were well established and reached 
a quasi-constant amplitude and structure.

We obtained the final simulation of the statistically steady state from a spin-up phase from a preliminary run with eightfold azimuthal symmetry. To guarantee that a statistically steady state {was} reached, we simulate{d} the preliminary run for over three viscous timescales with a final radial resolution of 217 Chebyshev grid points. We then lifted the azimuthal symmetry by running the simulation for an additional 0.4 viscous times in the entire azimuthal domain to be sure that a statistically steady state {was} reached. The final simulation use{d} 1024 grid points in longitude and 512 grid points in latitude, resolving spherical harmonic degrees until $\ell \leq 341$. After this equilibrated state has been reached, we continue{d} the simulation for another 0.6 viscous timescales and use{d} 416 realizations from this interval for the subsequent statistical analysis of the steady-state energy transfer.

\begin{figure*}
    \centering
    
    \includegraphics[width=\textwidth]{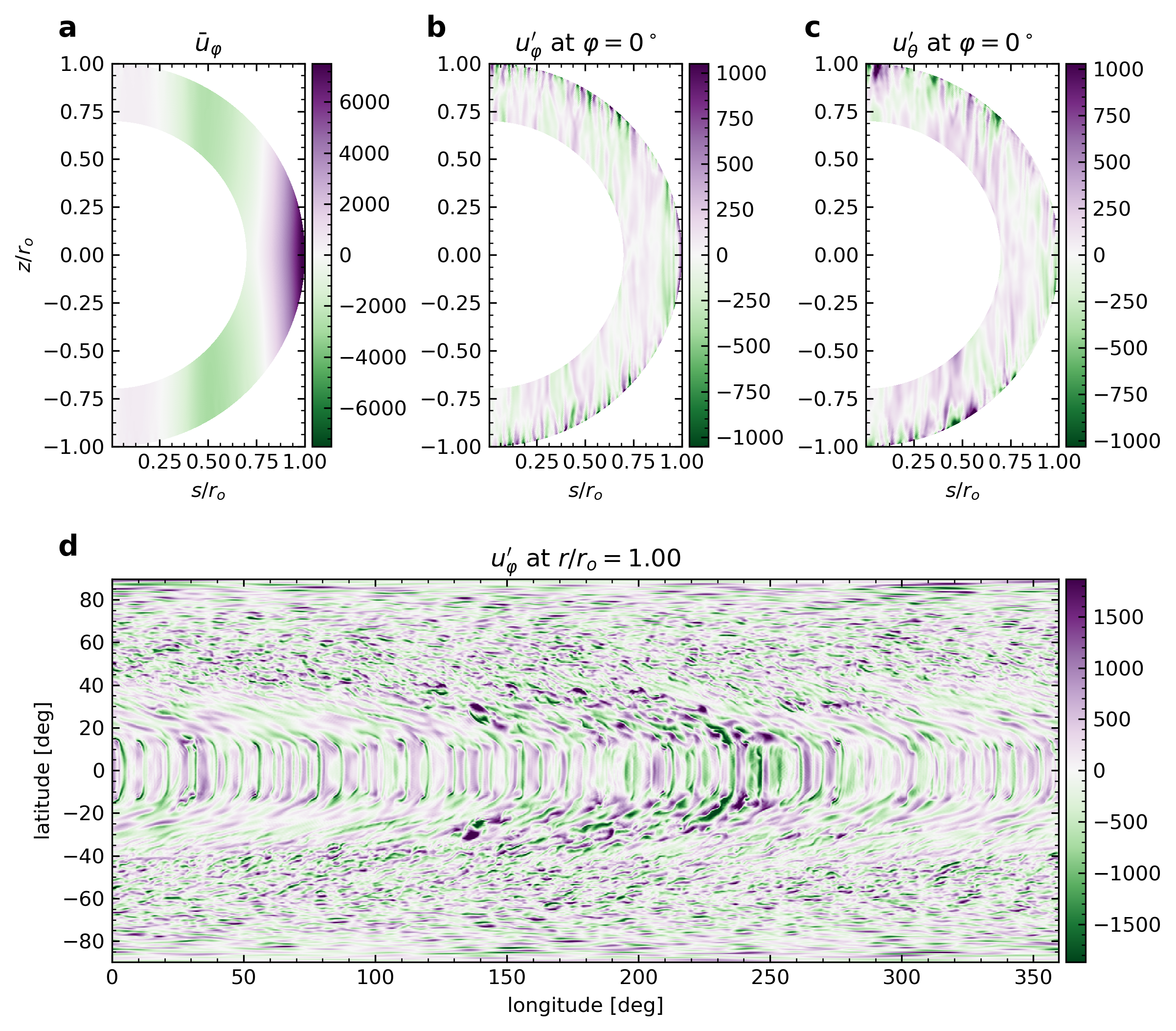}

    \caption{Typical simulation snapshots in the statistically steady state. Panel (a) shows the jet profile, $\bar u_\varphi$. Panels (b) and (c) show the fluctuations around the zonal mean of the zonal and meridional winds, $u_\varphi'$ and $u_\theta'$, at the central meridian. Panel (d) shows $u_\varphi'$ at the top boundary. Panels (b)-(d) are saturated at half of the maximum value in the domain shown. 
    }
    \label{figExampleSnapshots}
\end{figure*}

\begin{figure*}
    \centering
    \includegraphics[width=\textwidth]{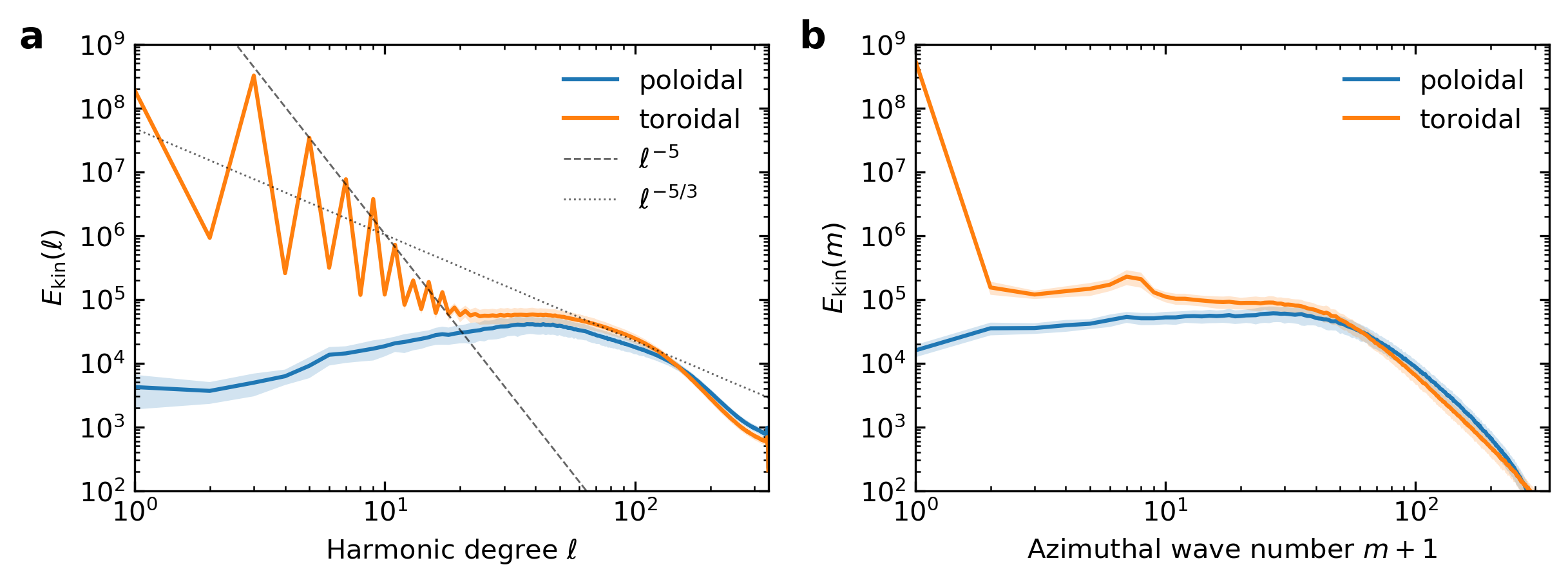}
    
    \caption{Kinetic energy spectra time-averaged over an interval that is part of the statistically steady state of our simulation as a function of harmonic degree (a) and azimuthal order (b). Both panels show separate spectra for the toroidal and poloidal flow components. {See Sect.~\ref{secGeostrophic} for a brief discussion of the possible power laws shown in panel (a).}}
    \label{figSpectra}
\end{figure*}

We show example snapshots from the statistically steady state of this run in Fig.~\ref{figExampleSnapshots}. The overall pattern of the flow is generally as expected for rapidly rotating convection in a spherical shell {\citep[e.g.,][]{Christensen2001}}. Due to the Taylor-Proudman theorem, the axisymmetric zonal flows $\bar u_\phi$ are predominantly geostrophic, and therefore invariant along the $z$ axis, as well as relatively time-invariant. The non-axisymmetric, turbulent small-scale convection $\vec{u}' = \vec{u}- \bar{\vec{u}}$ shows a complex spatial and time dependence with elongated features in the $z$-direction.

The jet profile is shown in Fig.~\ref{figExampleSnapshots}a. It consists of an equatorial prograde jet, flanked by a mid-latitude retrograde jet starting roughly at the tangent cylinder (TC; the cylindrical radius of the shell at the equatorial plane), which is again flanked by a weakly prograde polar jet. The jet profile is largely invariant along the $z$ axis, as expected from the Taylor-Proudman theorem. Small deviations due to thermal winds exist. The simulated jet profile is simpler than the actual profile on Jupiter or Saturn \citep[e.g.,][]{Vasavada2005,Genio2009}.  
Given that the simulations have to be run for multiple viscous timescales until a statistically steady state is reached, we restrict this study to one case that is somewhat simpler in nature but nevertheless includes the essential physics.

Figure~\ref{figSpectra} shows kinetic energy spectra, which we averaged over a time interval from the statistically steady state. For the spectra, the flow was d{i}vided into toroidal and poloidal components \citep[e.g.,][]{Christensen2015}. Toroidal flows are flows with radial vorticity and poloidal flows are flows with horizontal divergence including flows driven by convection. The spectra were computed as a function of harmonic degree $\ell$ and azimuthal wave number $m$ according to \citet{Christensen2015}. The toroidal kinetic energy for $m=0$ is due to the jets, while the poloidal kinetic energy for $m=0$ is associated with meridional flows. The zigzag shape of the toroidal kinetic energy in Fig.~\ref{figSpectra}a is due to the hemispheric symmetry of the zonal winds{, because the $m=0$ component dominates $E_{\rm kin}(\ell)$ for small $\ell$. The peak in the poloidal kinetic energy reflects the convective driving scale, which depends in a complex way on the system parameters and the geometry.}

In the following, we distinguish different characteristic azimuthal scales with different properties in the fluid motions based on Figs.~\ref{figExampleSnapshots} and~\ref{figSpectra} as well as based on our results (Sect.~\ref{secResults}). The flows consist of a strong zonal flow and of a weaker meridional flow (both at the jet scale, $m=0$), of large-scale eddies ($0<m  \lesssim 20$), of small-scale convection ($20\lesssim m \lesssim100$), and of flows at the classical dissipation scales ($m\gtrsim 100$). Toroidal {zonal} flows are dominant at large scales, while poloidal and toroidal flows have a similar amplitude at the convective and classical dissipation scales (see Fig.~\ref{figSpectra}).

\begin{figure}
    \centering
    \includegraphics[width=\linewidth]{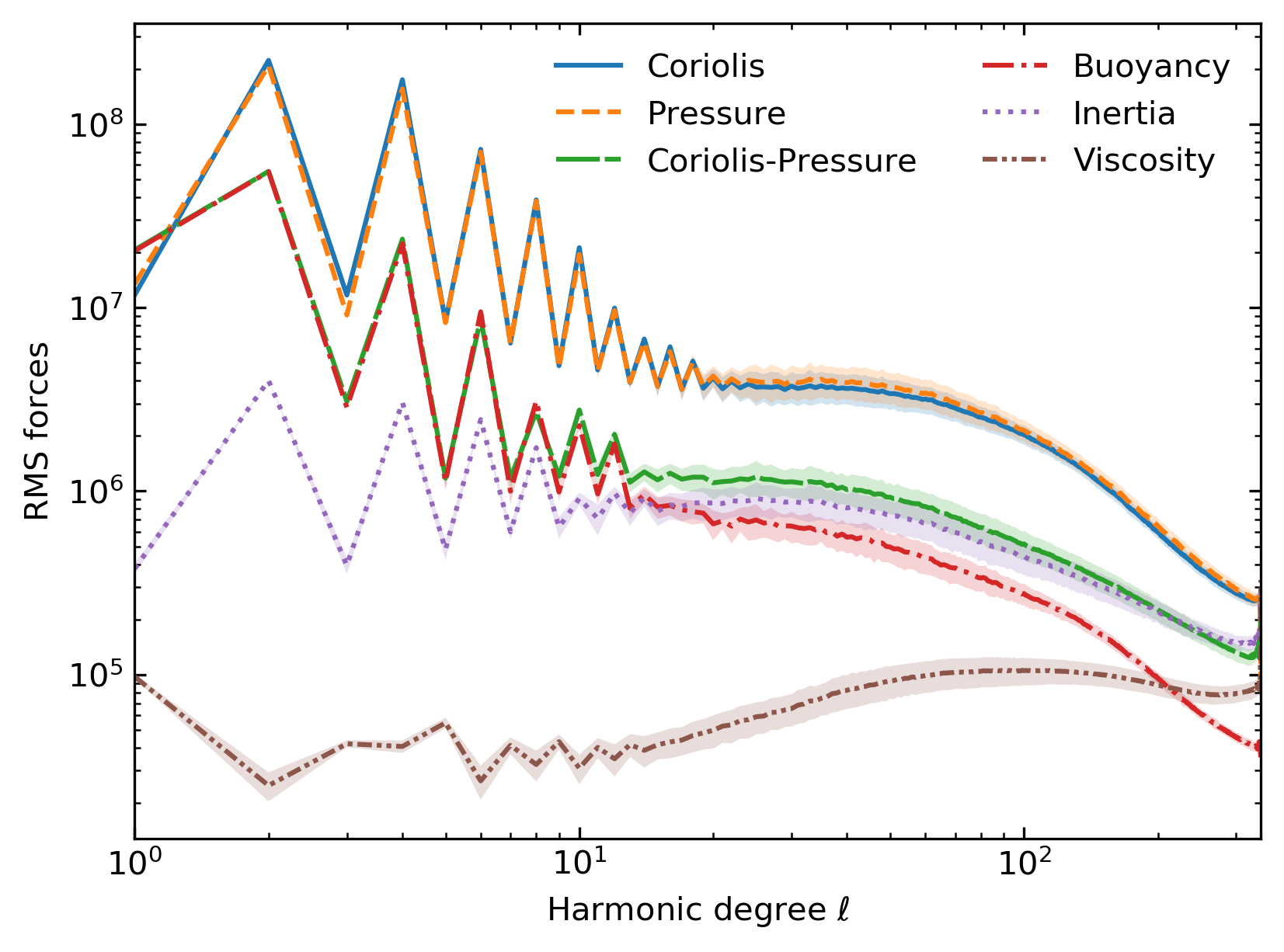}
    \caption{Square root of the shell-averaged squared forces in spectral space as a function of harmonic degree $\ell$ (also known as r.m.s. force spectra). The Coriolis force is dominating at all scales as is typical for rapidly rotating spherical-shell {convection}.}
    \label{figRMSforces}
\end{figure}

{The simulated flows and in particular the zonal winds are predominantly geostrophic, that is, variations in the directions of the rotation axis are minimal. Geostrophy is enforced by a primary force balance between Coriolis force and pressure gradient. 
This is clearly evident from Fig.~\ref{figRMSforces}, which shows the square root of the shell-averaged squared forces in spectral space (r.m.s. force spectra). The Coriolis force dominates at all scales, as it is typical for spherical-shell convection at small Ekman numbers.

Recently, the idea of a zonostrophic regime has been discussed in the context of jet formation in two-dimensional forced turbulence {on a $\beta-$plane or rotating spherical surface} \citep[e.g.,][]{Galperin2019}{, in shallow global climate models \citep[][]{Cabanes2020}, and in an experimental study of forced rotating turbulence \citep{Lemasquerier2022}. The zonostrophic regime is characterized by strong jets, which develop between the (large) Rhines scale and the (sufficiently smaller) Pelinovsky-Vallis-Maltrud scale \citep[][]{Galperin2006,Sukoriansky2007}. In the corresponding spectral region, the kinetic energy spectrum is dominated by zonal flows and shows a characteristic $\ell^{-5}$ scaling.
Interestingly, an approximate $\ell^{-5}$ scaling can be seen in our solution for $\ell<20$ (see Fig.~\ref{figSpectra}a). The scale $\ell=20$ marks the scale where buoyancy starts to dominate inertia in the r.m.s. force spectra (see Fig.~\ref{figRMSforces}), as expected for thermal Rossby waves \citep[e.g.,][]{Roberts1968,Busse1970}. This is consistent with the interpretation that the $\ell^{-5}$ range is dominated by Rossby waves \citep[][]{Sukoriansky2007}.

We therefore estimated the Rhines scale, $\ell_{\rm Rh}$, and the Pelinovsky-Vallis-Maltrud scale, $\ell_\beta$, and find
\begin{align}
    \ell_{\rm Rh} &= \sqrt{\frac{\beta}{2U_{\rm rms}}} = 4.7, \\
    \ell_\beta &\approx 30,
\end{align}
where we used $\beta=\Omega/R=E^{-1}/r_o$ \citep[see also][]{Galperin2019} and we estimated $\ell_\beta$ as the cross-over scale, where an average Rossby-wave period $\tau_{RW}=2\pi\ell/\Omega$ equals the turnover timescale $\tau_{t}=2\pi (U_\ell\, \ell/R)^{-1}$, where $U_\ell=\sqrt{2 \,\ell \, E_{{\rm kin},m \neq 0} / M}$ \citep[similar to ][]{Rhines1975} and $M$ is the total mass of the shell. We therefore find a zonostrophy index of $\ell_\beta / \ell_{\rm Rh}=6.4$, which implies that the $\ell^{-5}$ spectral range is sufficiently large for the zonostrophic regime \citep[e.g.,][]{Sukoriansky2007}.

We however also find differences to the zonostrophic regime. Firstly, the estimated $\ell_{\beta}\approx30$ is very close to the forcing scale, which is around $\ell_F\approx 40$. Strictly speaking, there has to be a minimum scale separation of at least a factor of two between the forcing scale and $\ell_\beta$ to permit an isotropic inverse cascade to develop \citep[e.g.,][]{Galperin2006,Sukoriansky2007,Galperin2019}. From zonostrophic theory, one might therefore expect that the inverse cascade or the characteristic $\ell^{-5}$ scaling do not show clearly in our case. On the other hand, it is possible that the forcing scale does not matter much for the development of the $\ell^{-5}$ scaling, as was also speculated by \citet[][]{Lemasquerier2022} but which is in contradiction to a statement by \citet[][]{Galperin2006}.}

Further research is necessary to firmly conclude on the $\ell^{-5}$ power law scaling of jets {in deep-shell convection}, in particular an analysis of spectral scaling and spectral transfer in harmonic degree, ideally in{cluding} a suite of {simulations}  spanning a larger range of parameters and with a number of jets inside the TC. At small scales, it is possible to fit an $\ell^{-5/3}$ slope to our spectra, but this slope is not unique due to the curved shape of the spectrum.

}

\section{Kinetic energy transfer between azimuthal wave numbers in a spherical shell}

\label{secKinEnBalanceTransfer}

\subsection{Scale-by-scale balance of kinetic energy}
\label{secbalance}

We here derive the expressions for the balance and flux of kinetic energy at azimuthal wave numbers $m$ using the Cartesian formulation of \citet{Alexakis2018} as a guideline. To do this, we perform a Fourier transform in $\varphi$ of all variables using the normalization of \citet[][]{Lago2021}. The kinetic energy is 
\begin{align}
 E_{\rm kin} &= \int_V \id^3\br \, \frac{1}{2}\rho  |\vec u|^2 =  \sum_{m\geq0} E_{\rm{kin}}(m), \label{eqEkinTot}
 \end{align}
where $V$ is the full spherical shell volume and
\begin{align}
 E_{{\rm kin}}(m) &= \int_V \id^3\br \, \frac{1}{2}\rho \,  |\vec u^{(m)}|^2
 \label{eqEkinm}
\end{align}
is the kinetic energy at azimuthal wave number $m$. Here, $\vec u^{(m)}$ is the flow in physical space filtered in such a way as to retain only the $m$-th azimuthal Fourier component. 
The equations in this section are valid both for decomposing into spherical and cylindrical coordinates. By taking the time derivative of the kinetic energy (Eqs.~\eqref{eqEkinTot} and~\eqref{eqEkinm}) and inserting Eq.~\eqref{eqNSE}, we obtain the kinetic energy balance
\begin{align}
 \partial_t  E_{\rm{kin}}(m) &= T +F_{P} + F_{C} + F_{\rm{Buoy}} - D, \label{eqEbalanceScale}
\end{align}
where the individual contributing terms are due to nonlinear advection ($T$), work done by pressure ($F_P$), by buoyancy ($F_{\rm{Buoy}}$), by the Coriolis force ($F_C$), and by viscous dissipation ($D$),
\begin{align}
T(m) &= - \int_V \id^3\br \, \rho \, \langle \vec u^{(m)}, (\vec{u}\cdot\vec{\nabla}) {\vec u} \rangle,  \label{eqTm}\\
F_{P}(m) &= - \int_V \id^3\br \, \rho  \,\langle \vec u^{(m)}, \vec{\nabla}\left({\dfrac{p}{\tilde{\rho}}}\right) \rangle=0,  \\
F_{C}(m) &=  - \int_V \id^3\br \, \rho  \, \langle \vec u^{(m)}, \dfrac{2}{E} (\vec{e_z}\times\vec{u})  \rangle  = 0,\\
F_{\rm{Buoy}}(m)&= \int_V \id^3\br \, \rho  \,\langle \vec u^{(m)}, \dfrac{Ra}{Pr}\tilde{g} \,S\,\vec{e_r}\rangle, \label{eqFBuoy}  \\
D(m) &= \int_V \id^3\br \, \rho  \,\langle \vec u^{(m)}, \dfrac{1}{\tilde{\rho}} (\vec{\nabla}\cdot \mathcal{S}) \rangle . \label{eqDm}
\end{align}
We here use $\langle \bm{a},\bm{b} \rangle$ for the vector product $\bm{a} \cdot \bm{b}$ and all expressions are functions of time. The Coriolis force merely transfers energy between the $s$ and $\varphi$ flow components and therefore does no net work. Consequently, the integrand of $F_C(m)$ is zero at every location. 
The work done by the pressure force is zero at any $m$ because we integrate over the entire shell.

In addition to the global energy balance, we also consider the distribution of the local contributions to Eq.~\eqref{eqEbalanceScale} in the meridional plane $(r,\theta)$. To this end, we replace the volume integrals in Eqs.~\eqref{eqTm} to~\eqref{eqDm} by a simple integration over the $\varphi$-coordinate for each $(r,\theta)$ and denote the result with the same letter for simplicity. 
More formally, for any of the quantities $Q=E_{\rm kin},T,F_{\rm Buoy}$, we write
\begin{align}
 Q(m) &= \int r\, \id r \, \id \theta \,  Q (m,r,\theta).
 \label{eqQq}
\end{align}

\subsection{Kinetic energy transfer from scale to scale}
\label{sectransfer}

Only the nonlinear advection term $T(m)$ is able to transport kinetic energy from one spectral scale to another. {We evaluate this scale-to-scale transfer of kinetic energy by $T(m,m')$, which is a scale-by-scale version of $T(m)$ and given by
\begin{align}
T(m,m') &= \int_V \id^3\br \, \rho  \, \langle \vec u^{(m)}, (\vec{u}\cdot\vec{\nabla}) \vec u^{(m')} \rangle.  \label{eqTmm}
\end{align}
The relationship between $T(m)$ and $T(m,m')$ is given by
\begin{align}
T(m) &= \sum_{m'\geq 0} T(m,m').
\end{align}
}
The energy transfer function, $T(m,m')$, describes how nonlinear advection transports kinetic energy from scale $m'$ to scale $m$. 
Positive $T(m,m')$ means a transfer of energy from wave number {$m'$ to $m$}, which is also known as forward transfer if it is to smaller scales ($m>m')$ and upscale transfer if it is to larger scales ($m<m'$). We note that different choices for the sign of $T$ appear across the literature \citep[e.g.,][]{Rubio2014,Favier2014,Alexakis2018}. Due to energy conservation, we have $T(m,m') = - T(m',m)$.

\subsection{Energy transfer to the jets}
\label{secReynolds}

The $m=0$ component of the kinetic energy represents mainly the zonal winds, since the meridional circulation, the second axisymmetric flow contribution, is orders of magnitude smaller. 
Since the axisymmetric zonal winds are a toroidal flow, they do not experience a driving by buoyancy. Therefore, advection mainly feeds energy into the jets from the eddies ($m>0)$. Using simple orthogonality relations and a partial integration, one can show that
\begin{align}
T(0,m') &=   \int r^2 \sin \theta\, \id r \, \id\theta  \, 2\pi \,(\tilde \rho\,\overline{ u_i^{(m')}  \vec{u}^{(-m')} } \cdot\vec{\nabla}) u_i^{(m=0)}, \label{eqTm0mprime}
\end{align}
where we have used the summation convention over $i$ and the overbar denotes an azimuthal average. As a consequence, the total energy transfer to the $m=0$ flow is
\begin{align}
T(m=0) &=  \int r^2 \sin \theta\, \id r \, \id\theta \, 2\pi \, ( \vec{R_i} \cdot\vec{\nabla}) u_i^{(m=0)}, \label{eqTm0Reynolds}
\end{align}
where 
\begin{align}
R_{ij} &= (\vec{R_i})_j = \tilde \rho\, \overline{ u_i'  u_j' } = \sum_{m' > 0} R_{ij,m'}
\end{align}
is the Reynolds stress from the entire flow field. Here, the quantity
\begin{align}
R_{ij,m'} &= \tilde \rho\, \overline{ u_i^{(m')}  u_j^{(-m')} }
\end{align}
is the Reynolds stress, which is responsible for the energy transfer from a fluid scale $m'>0$ to the jet scale $m=0$ (see Eq.~\eqref{eqTm0mprime}). The energy transfer to the jets is entirely due to Reynolds stresses. The value $T(0,0)$ describes the generation and redistribution of zonal flow by meridional circulation. This is very small locally and averages to zero over the whole convective shell.

\subsection{Scale-to-scale flux of kinetic energy}
\label{secflux}

In addition to the scale-by-scale balance of kinetic energy, it is interesting to consider the flux of kinetic energy between different ranges of scales, which is mediated by the advection term $T(m)$. This can be obtained by a cumulative sum over $m$,
\begin{align}
\partial_t  \sum_{\tilde m\leq m} E_{\rm{kin}}(\tilde m) &= - \Pi(m) +  \epsilon_{\rm{Buoy}}^{\leq m}   - \epsilon_{D}^{\leq m},  \label{eqEbalanceFlux}\\
\Pi(m) &= - \sum_{\tilde m\leq m} T(\tilde m)= \sum_{\tilde m > m} T(\tilde m), \label{eqPi}\\
\epsilon_{\rm{Buoy}}^{\leq m} &= \sum_{\tilde m\leq m} F_{\rm{Buoy}}(\tilde m), \quad \epsilon_{D}^{\leq m} = \sum_{\tilde m\leq m} D(\tilde m),
\end{align}
where we used $T(m,m') = - T(m',m)$ in Eq.~\eqref{eqPi}. 
The scale-to-scale flux $\Pi(m)$ is the kinetic energy received by scales $\tilde m > m$ from scales $\tilde m \leq m$ by advection and thus characterizes the energy flux across scale $m$. Positive $\Pi(m)$ means transfer across $m$ from larger to smaller scales, which is also known as forward flux, and negative $\Pi(m)$ is transfer across $m$ to larger scales, which is also known as upscale (or inverse) flux.

In total, the nonlinear advection term therefore acts neither as a source nor as a sink of kinetic energy because $\lim_{m\to\infty} \Pi(m) = 0$. 
The total kinetic energy balance for the entire shell reads
\begin{align}
    \partial_t \sum_{m\geq 0} E_{\rm kin} = \sum_{m\geq 0}  (F_{\rm Buoy} - D),
\end{align}
and buoyancy is the global energy source and dissipation the global energy sink.

When the simulation reaches a statistically steady state, the temporal mean of the kinetic energy does not change any more to a significant extent. From here on, we consider temporal averages of the above terms over sufficient time such that their values are close to their long-term means. The time derivative of the time-averaged kinetic energy is thus very small and can be neglected, $\partial_t E_{kin}(m)=0$, and the balances \eqref{eqEbalanceScale} and \eqref{eqEbalanceFlux} become
\begin{align}
     T + F_{\rm{Buoy}} - D  &= 0, \label{eqEbalanceScaleSS} \\
     -\Pi + \epsilon_{F_{\rm{Buoy}}}^{\leq m}   - \epsilon_{D}^{\leq m} &= 0 \label{eqEbalanceFluxSS}.
\end{align}
Equations~\eqref{eqEbalanceScaleSS} {and~\eqref{eqEbalanceFluxSS}} show that the advection term balances the source and sink terms of the kinetic energy at each $m$. It transports kinetic energy away from the buoyantly forced scales to the dissipation scales.

\section{Results for the statistically steady state}
\label{secResults}

\subsection{Scale-to-scale flux: Upscale flux at large scales, forward flux at small scales}

\label{secScaleByScaleBalance}

\begin{figure*}
    \centering

    \includegraphics[height=0.35\textwidth]{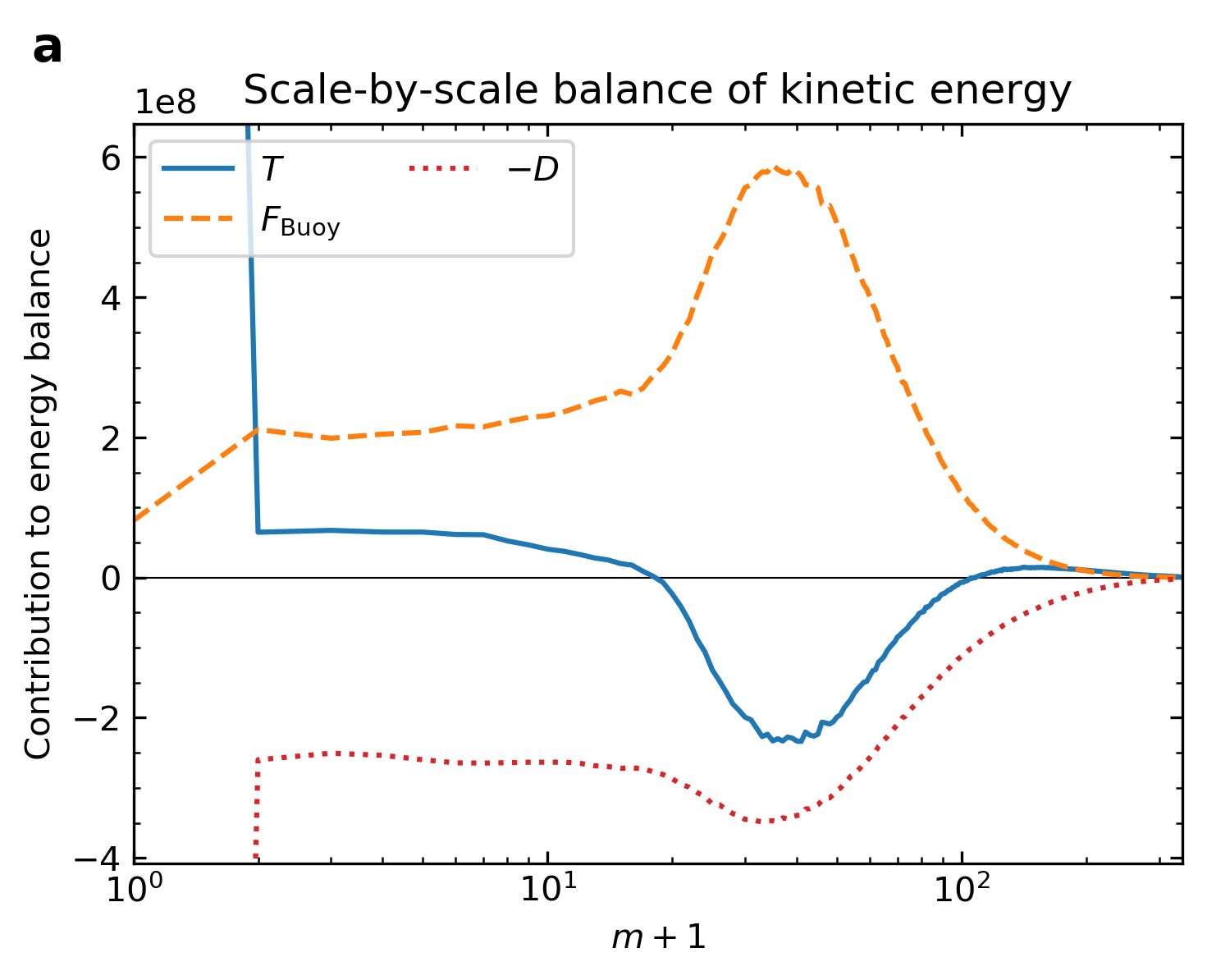}
    \includegraphics[height=0.35\textwidth]{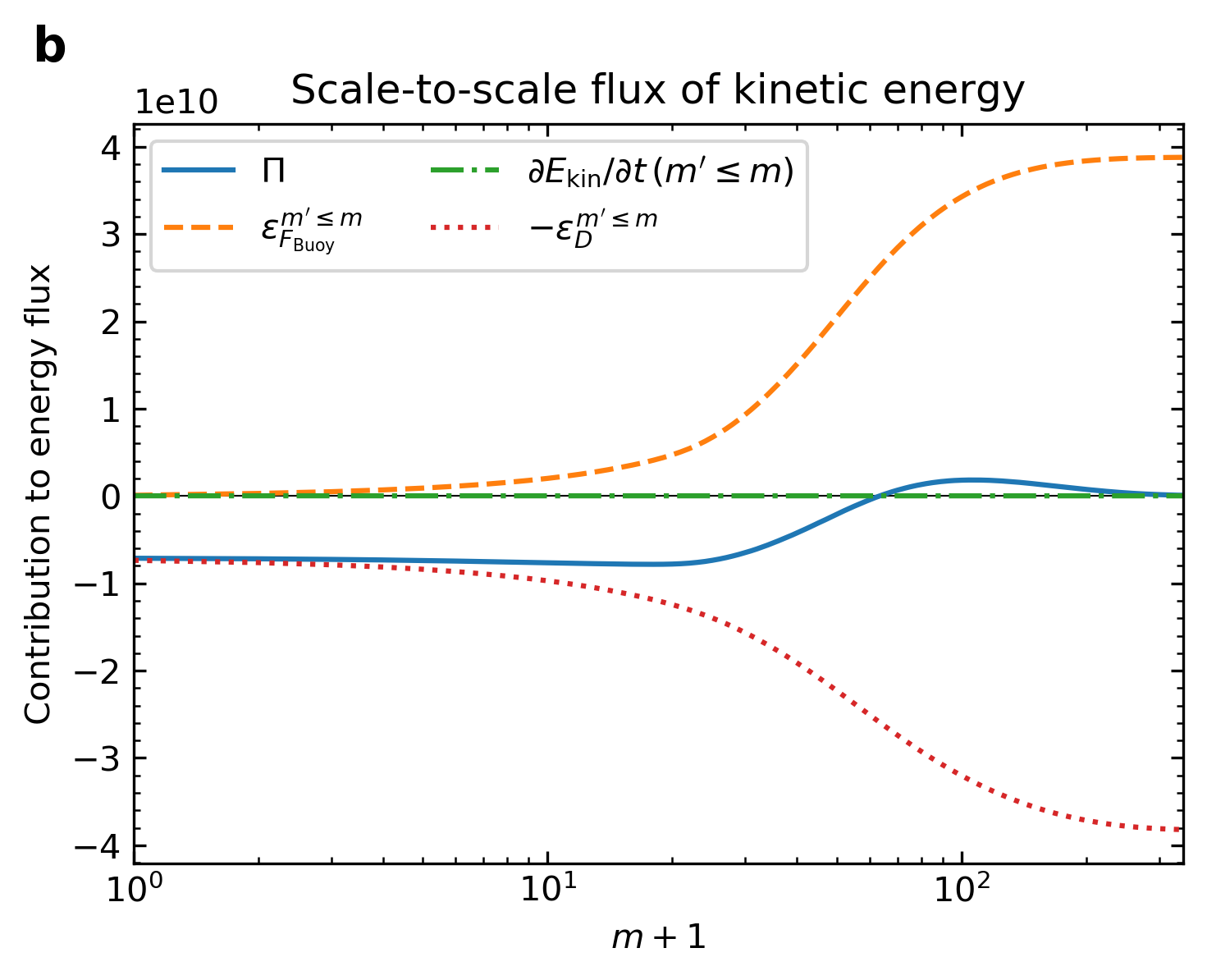}

    \caption{Scale-by-scale balance (a) and scale-to-scale flux (b) of kinetic energy in the statistically steady state. The values $T(0)\approx D(0) \approx 7\times10^9$ were excluded from the plotting range in panel (a) for better visibility.}
    \label{figTPi}
\end{figure*}

As a first result, we show the scale-by-scale spectral balance of kinetic energy in Fig.~\ref{figTPi}a. Buoyancy work drives the kinetic energy at all scales, in agreement with Cartesian simulations of Rayleigh-Bénard convection \citep[e.g.,][Fig. 3]{Verma2019}. The dominant contribution (79 \%) to the convective driving comes from the small convective scales. 
The amplitude of the convective driving drops steeply for larger wave numbers but retains a sizable value for the smallest wave numbers.  At all scales except $m=0$ and $m\gtrsim 200$, buoyancy is the strongest driver of the motions. The small buoyant driving at $m=0$ injects energy into the meridional flow.

From the small convective scales, nonlinear advection transfers kinetic energy away to other scales ($T$ is negative). It is transferred to the jets and to large-scale eddies, as well as to the classical dissipation scales, where $T$ is positive. 
As a consequence, there is a net upscale flux of kinetic energy at larger scales ($\Pi<0$ at $m<60$). The scale-to-scale flux is relatively constant across the largest scales ($m<20$), so that the constant scale-to-scale flux criterion for an inverse cascade is satisfied \citep[e.g.,][]{Young2017,Alexakis2018}.

At smaller scales, there is a net forward flux of kinetic energy to ever smaller scales with $\Pi$ positive and nearly constant for $80\lesssim m \lesssim 150$. Such an approximately constant forward flux might similarly be associated with the inertial range of a forward cascade.

Viscous dissipation extracts energy from all scales as expected. The small convective scales experience the strongest dissipation (46 \%), while the classical dissipation scales give rise to only a fraction of the dissipation (18 \%). Dissipation at the jet scale contributes 26 \% to the total dissipation. {Since we use free-slip boundary conditions, there is no Ekman boundary layer and the dissipation in general takes place in the bulk of the domain. There is some tendency for increased dissipation toward the outer boundary of the shell, likely because the convective motions become stronger in that region.} The strong role of dissipation at the convective and large scales ($m<100$) is most likely due to the use of a turbulent viscosity. The viscosity of actual planetary interiors is such that the Ekman number is much smaller than in our study, which results in viscosity acting only on the smallest scales. In an actual planet, nonlinear advection to these unresolved scales is likely to play the role that the turbulent viscosity plays in our simulations.

\subsection{Results for the scale-to-scale transfer of kinetic energy}

\subsubsection{Upscale transfer of kinetic energy into jets occurs directly from most buoyant scales via Reynolds stresses}
\label{secUpscaleReynoldsNotACascade}

\begin{figure*}
    \centering

    \includegraphics[height=0.5\textwidth]{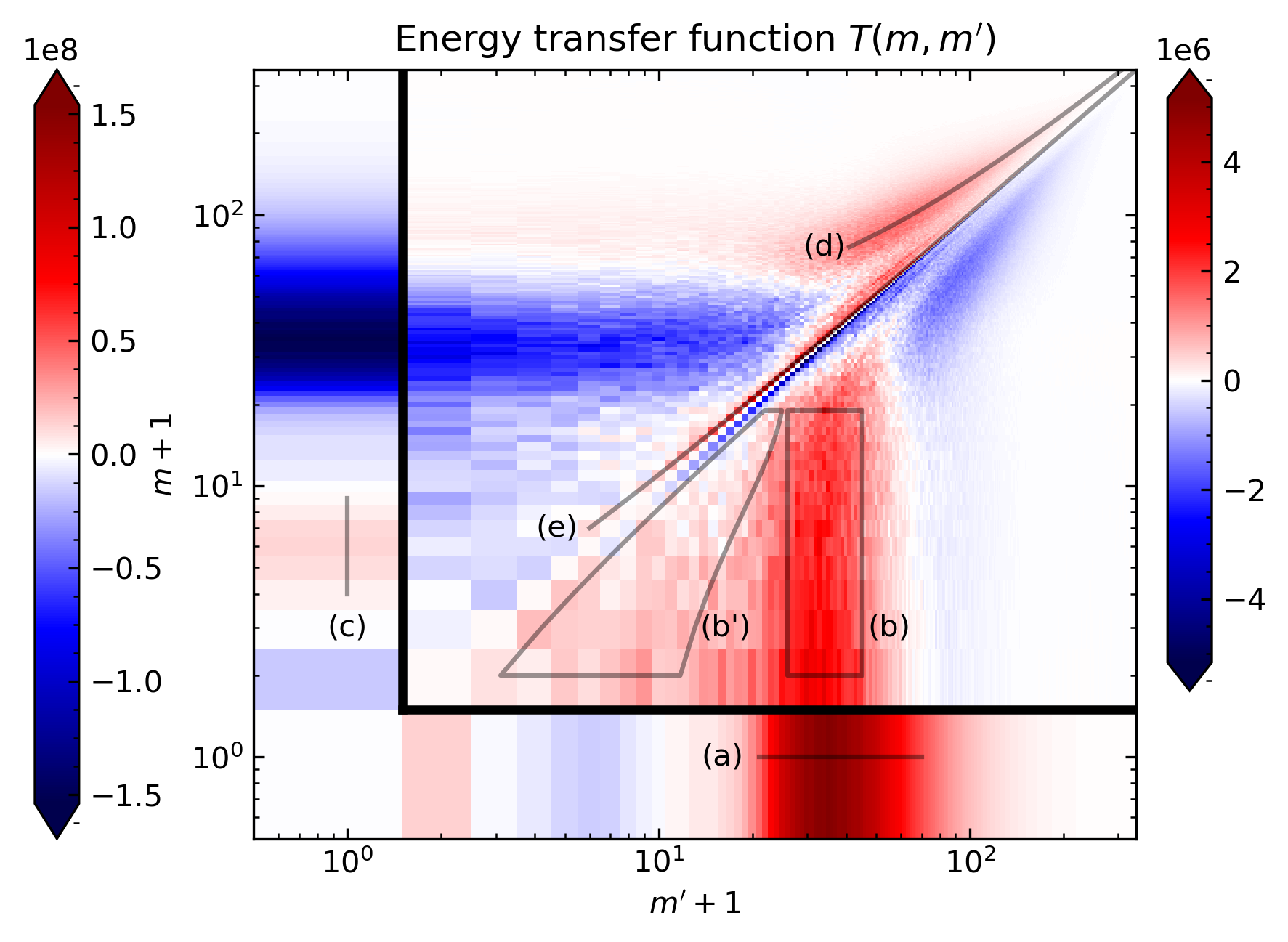}

    \caption{Scale-to-scale transfer of kinetic energy, $T(m,m')$, in the statistically steady state. Note that the figure has logarithmic axes and two colorbars; the left one indicates values for $m=0$ or $m'
    =0$ and the right one for $m,m'>0$. A positive value of $T(m,m')$ shows that kinetic energy is transferred from wave number $m'$ to wave number $m$. We use the regions indicated by gray lines and labeled with letters (a) to (e) to refer to specific regions of the plot in the text and show the different spatial contributions to $T(m,m')$ for those regions in Fig.~\ref{figTrtheta}.}
    \label{figTmm}
\end{figure*}

\begin{figure*}
    \centering

    \includegraphics[height=0.4\textwidth]{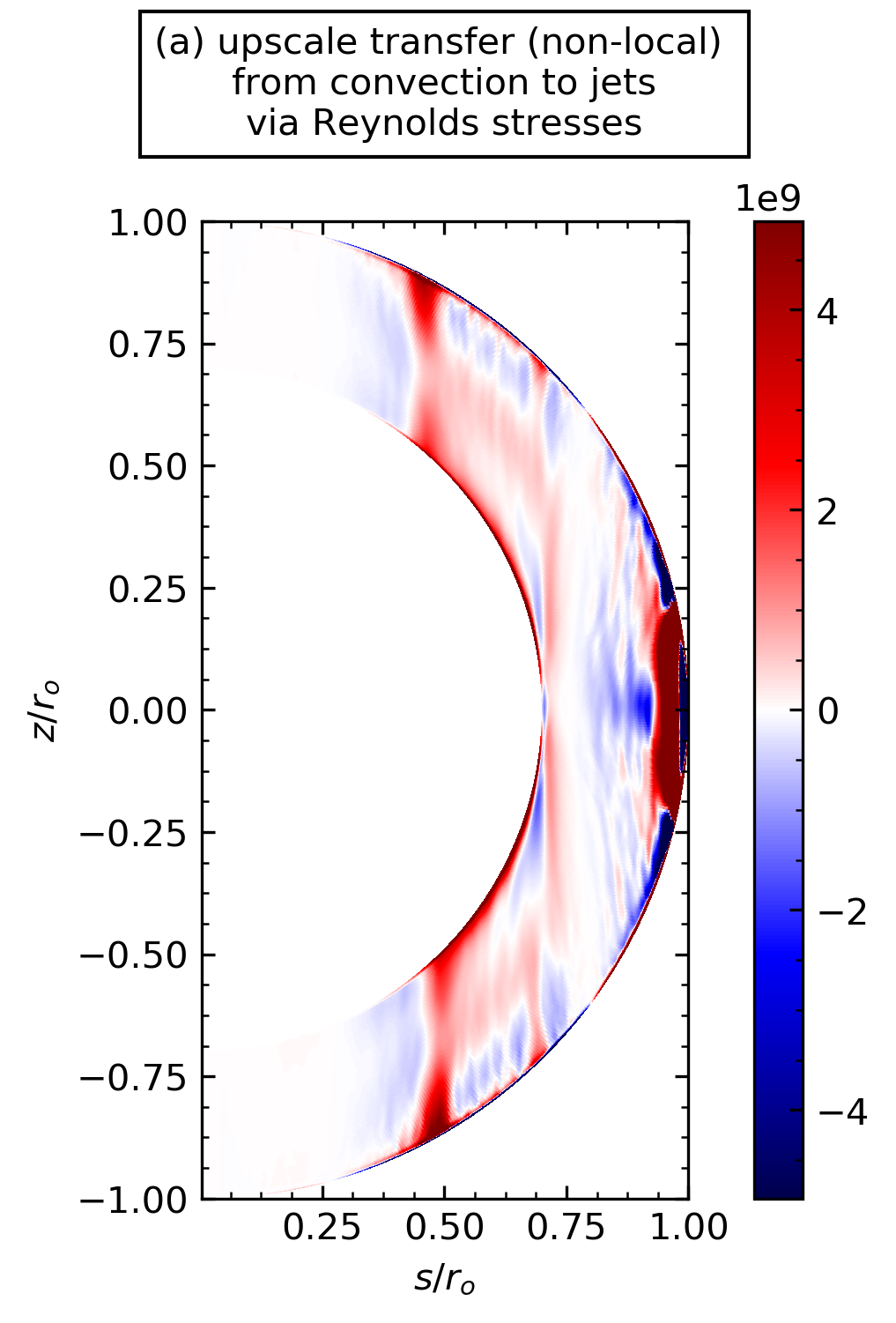}
    \includegraphics[height=0.4\textwidth]{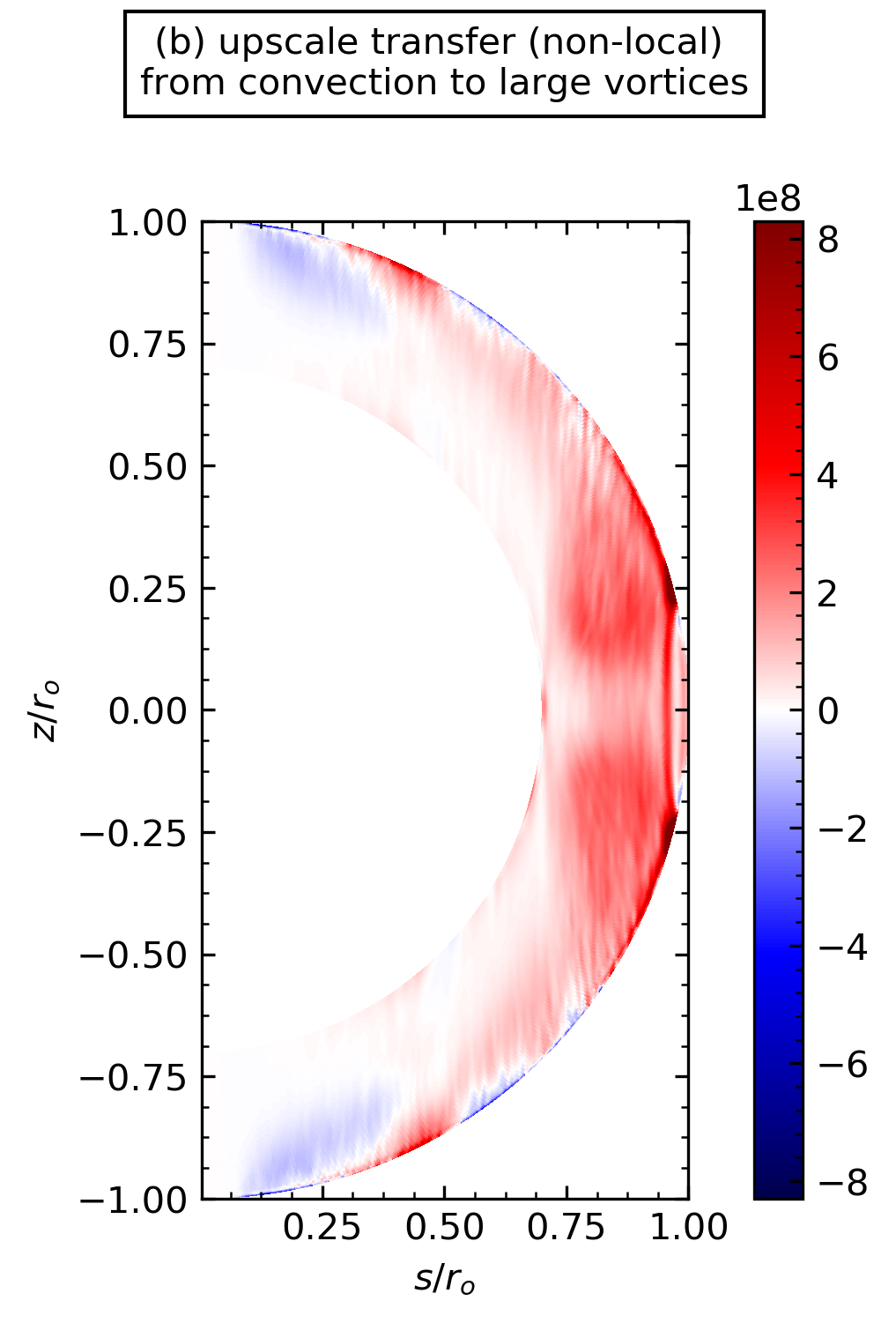}
    \includegraphics[height=0.4\textwidth]{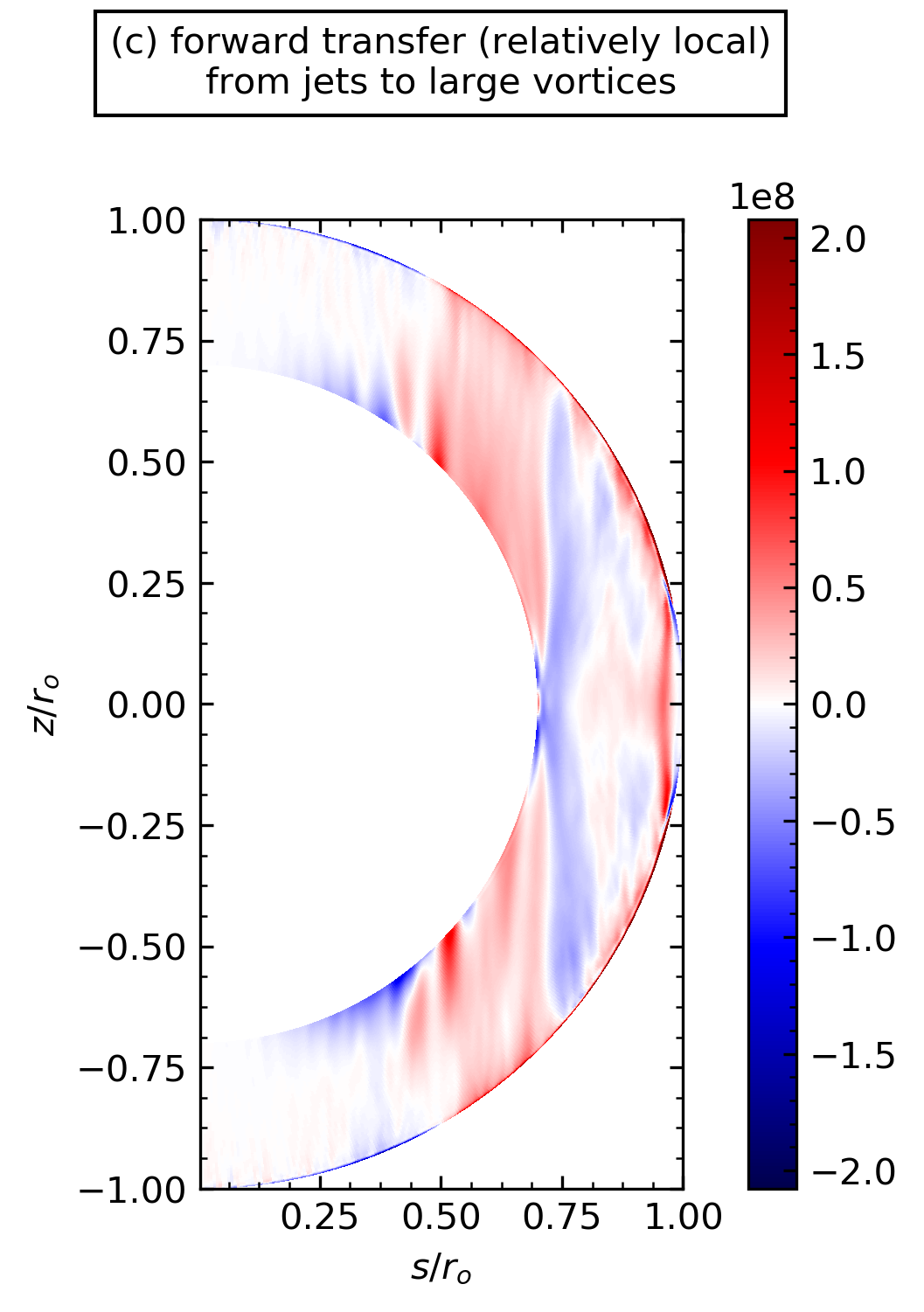}
    
    \includegraphics[height=0.4\textwidth]{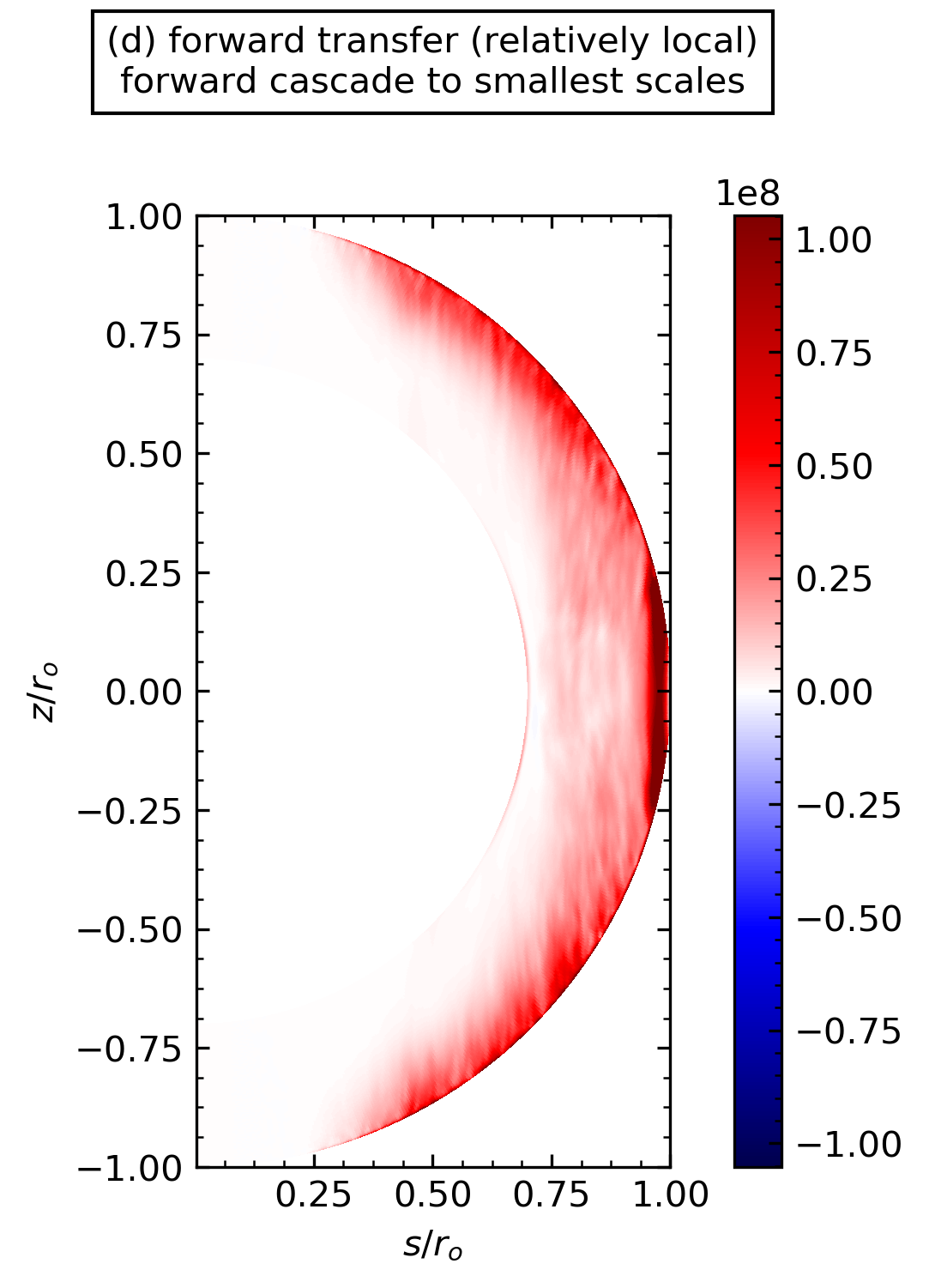}
    \includegraphics[height=0.4\textwidth]{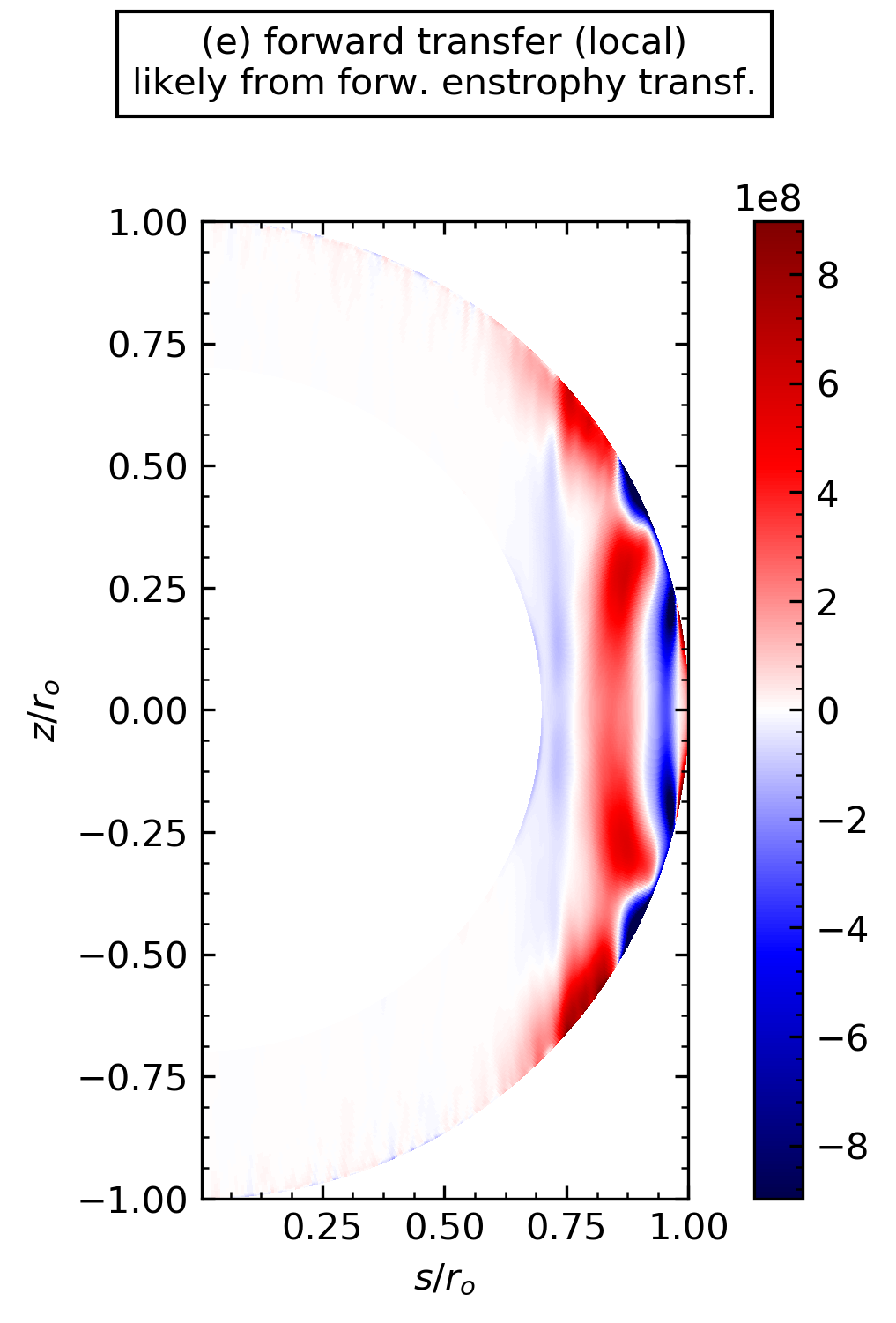}

    \caption{Spatial contribution, $T_i(r,\theta)$, to the energy transfer function, $T(m,m')$, summed over different values of $m$ and $m'$ according to the masks $(i)=(a),(b),(c),(d),(e)$ shown in Fig.~\ref{figTmm}. The panels display results for the direct driving of the jets by the convective scales via upscale energy transfers and Reynolds stresses (a), the upscale transfer to large eddies (b), the feeding of large-scale vortices from the jet energy (c), the small-scale forward cascade (d), and the local forward transfer likely associated with a forward enstrophy transfer (e). All panels are saturated at three times their maximum value, except for panel (a), which is saturated at 18 times its maximum value. 
    }
    \label{figTrtheta}
\end{figure*}

We show results for the energy transfer function in Fig.~\ref{figTmm}. We highlight different features in the bi-dimensional wavenumber space $(m,m')$ that are indicative of different modes of energy transfer by gray lines or polygons and discuss their meaning below. Overall, we find the largest values for the energy transfer function for $m=0$ and $m'=0$. In regions that involve $m=0$ or $m'=0$ the amplitude of the transfer is roughly 30 times larger than elsewhere. From the large amplitude in the region labeled (a), it is clear that this is energy transfer into the jets, rather than into the much weaker meridional flow. This transfer is caused by statistical correlations of flow components (Reynolds stresses; see Sect.~\ref{secReynolds}). The energy transfer by Reynolds stresses arises predominantly from the small convective scales that experience the strongest convective driving (region (a) in Fig.~\ref{figTmm}). {This transfer is extremely nonlocal in wavenumber space. It takes place directly from the convectively forced scales to the jets and bypasses a transfer into scales of intermediate size, from which the energy could have been transferred to the jets in a later step. We thus find that the jets are in the statistically steady state directly driven by the small convective scales rather than by an inverse cascade. We further underpin this conclusion in Sect.~\ref{secResultsSpinup} by analyzing the evolution of the simulation until the statistically steady state.} The steady-state transfer of kinetic energy from the convective scales to the jet resembles to some extent the energy transfer in simulations of rapidly rotating Rayleigh-Bénard convection \citep[e.g.,][]{Rubio2014,Favier2014,Kunnen2016}. These authors, however, seem to associate this upscale transfer with an inverse cascade despite the non-locality of the transfers {in their simulations}.

Figure~\ref{figTrtheta} shows the spatially resolved contribution to the energy transfer function. Here, we have summed the local contribution $T(m,m',r,\theta)$ over all $(m,m')$ that are part of a spectral region labeled in Fig.~\ref{figTmm}. From Fig.~\ref{figTrtheta}a, we take that the injection of energy into the jets is rather localized at the centers of the prograde and retrograde jet flow.

\subsubsection{Nonlocal upscale transfer to large vortices}
\label{secUpscaleTransferCascade}

For the transfer of energy to large-scale eddies, we observe two different types of behavior. Firstly, we find a region of upscale energy transfer from the scales with the largest buoyancy driving to comparatively larger scales (regions (b) and (b') in Fig.~\ref{figTmm}). This way, kinetic energy is transferred to larger scales approaching the jet scale. In this region, the more nonlocal transfers (region b) are stronger than the transfer in the more local region (b') by a factor of over 50, when summed over the entire region. The total transfer from region (b) is about seven times smaller than the transfer from region (a).

Region (b) has a small negative contribution to the upscale flux shown in Fig.~\ref{figTPi}b. The transfers are however quite nonlocal and can be understood as follows. The dominant convective scale of $m'\approx 35$ can be expected to have a strong nonlinear interaction with other scales of similar size, say $m''\approx 20-50$. The resulting third coupling partner then necessarily has the scale $m=m' \pm m''$. As rotational effects induce upscale transfers in addition to the usual forward transfers, this results in transfers to $m=m'-m''\approx 0-15$. 
This feature is therefore most likely a side effect of the transfer from the convective scales to the jet scale not always exactly matching $m=0$. The correlations of convective motions are not always perfect and may thus instead of feeding $m=0$ also feed $m=1,2,\ldots$ to a lesser extent. This possibility is supported by the maximum of the upscale transfer for each $m$ being always near the strongest convective scale, $m'\approx 35$. This scenario seems more plausible than interpreting region (b) as a signature of an inverse cascade because it explains the non-locality of the transfers. We therefore conclude that the upscale transfers in region (b) are likely not associated with an inverse cascade. 

\begin{figure}
    \centering

    \includegraphics[width=\linewidth]{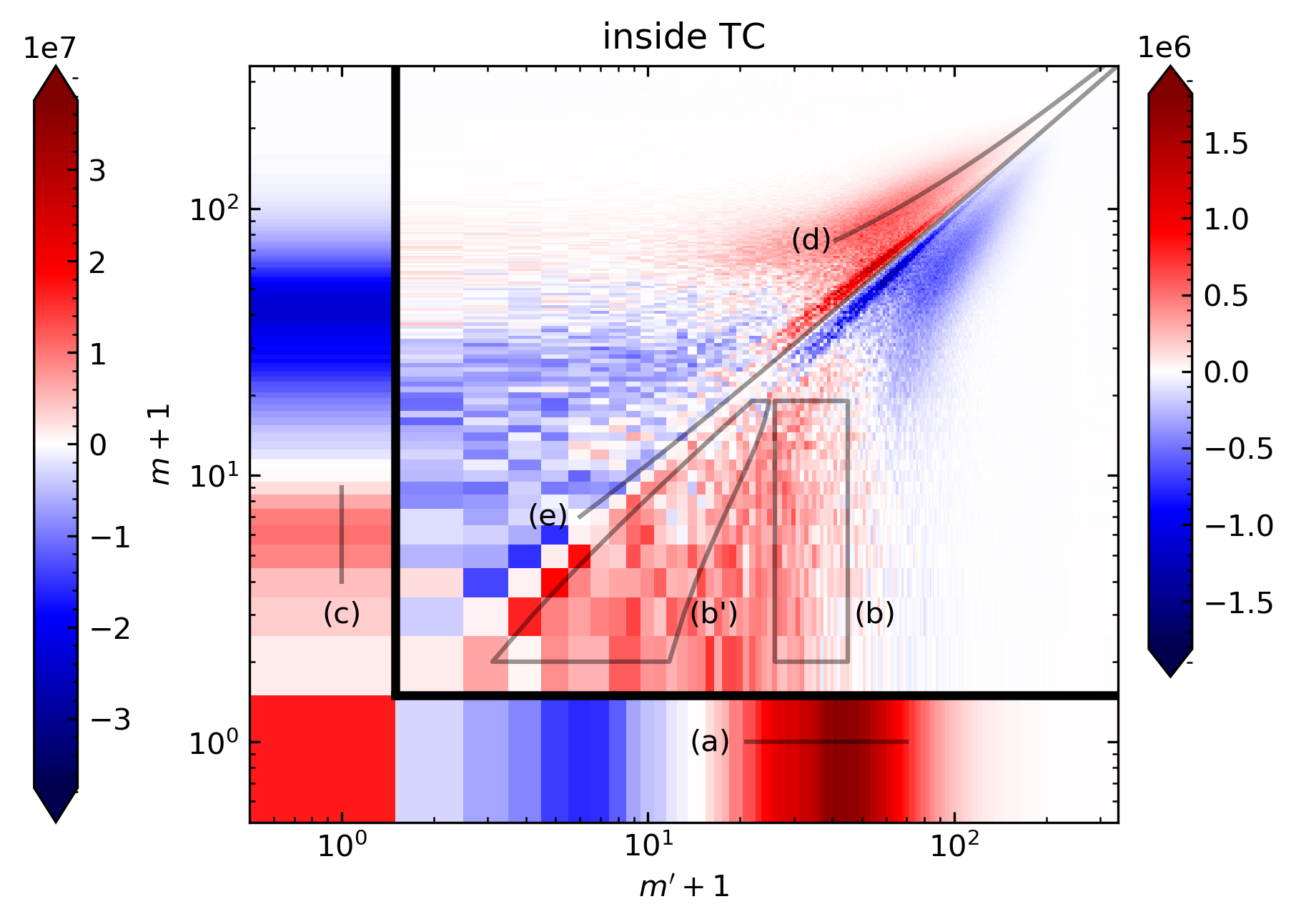}

    \caption{Energy transfer function, $T(m,m')$, in the statistically steady state, integrated over the region inside the TC $(s<r_i)$.}
    \label{figTmminsideTC}
\end{figure}

In region (b'), we find local positive transfers that could be a signature of an inverse cascade. In this region, however, the transfers are significantly weaker than in the neighboring region (b) and contribute barely to the upscale flux. We therefore do not interpret this feature as a signature of a global inverse cascade, which operates in the entire domain. The transfers in region (b') are predominantly caused by flows inside the TC. We therefore show the energy transfer function integrated only over the region inside the TC in Fig.~\ref{figTmminsideTC}. We find that the transfers in region (b') inside the TC are closer to the diagonal and therefore more local. Our results therefore indicate the possibility that large-scale vortices are driven by an inverse cascade inside the TC. The jets, however, are clearly also driven directly by the small convective scales also inside the TC (see region a in Fig.~\ref{figTmminsideTC}).

\subsubsection{Transfer from the jets to large-scale vortices: Jet instability as an alternative formation mechanism}

In addition to the upscale transfer, we find that large-scale vortices receive kinetic energy from the jets (see region (c) in Fig.~\ref{figTmm} and panel (c) in Fig.~\ref{figTrtheta}). This forward transfer from the jets is quite confined in $m$ compared to the rather broad upscale transfer region (b).

\begin{figure}
    \centering

     \includegraphics[width=0.7\linewidth]{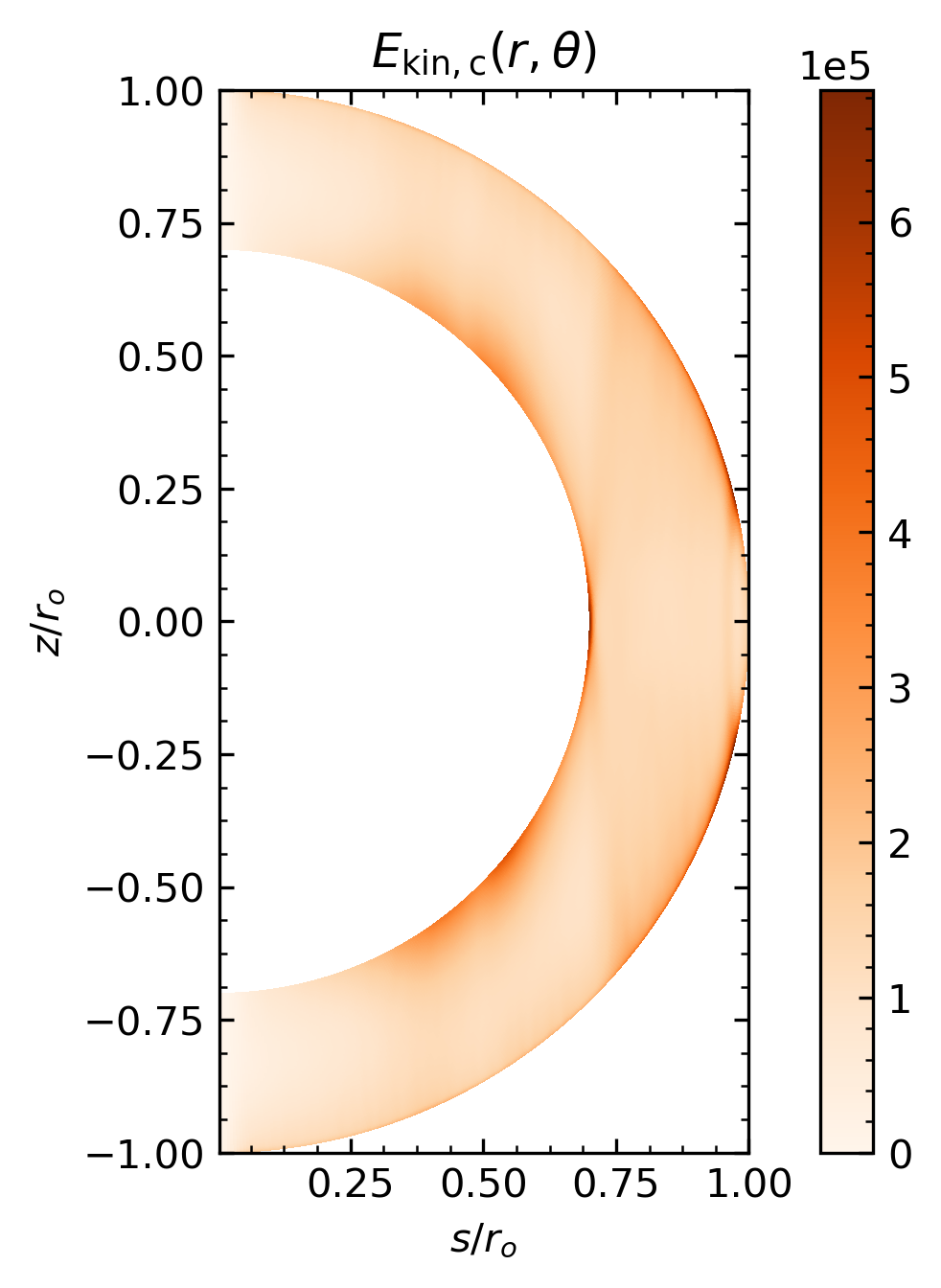}

    \caption{Spatial dependence of the contribution to the kinetic energy in the statistically steady state, $E_{\rm kin}$, for a sum over azimuthal wave numbers $m=3,\ldots,8$ (see label (c) in Fig.~\ref{figTmm}). See Eq.~\eqref{eqQq} for a definition of $E_{\rm kin}$.
    }
    \label{figEkin_mrtheta}
\end{figure}

The driving of large-scale vortices by a transfer from the jets is further supported by a comparison to the toroidal energy spectra. We find that there is a small peak in toroidal kinetic energy at those scales and centered around $m=5$ (see Fig.~\ref{figSpectra}b). The spatially resolved contribution to the kinetic energy, $E_{\rm kin} (m,r,\theta)$ further supports this conclusion. Figure~\ref{figEkin_mrtheta} shows that for $m=3,\ldots,8$, there is a noticeable excess power near the location of our retrograde jet at $s \approx 0.5$.

\begin{figure*}
    \centering
    \includegraphics[width=0.7\textwidth]{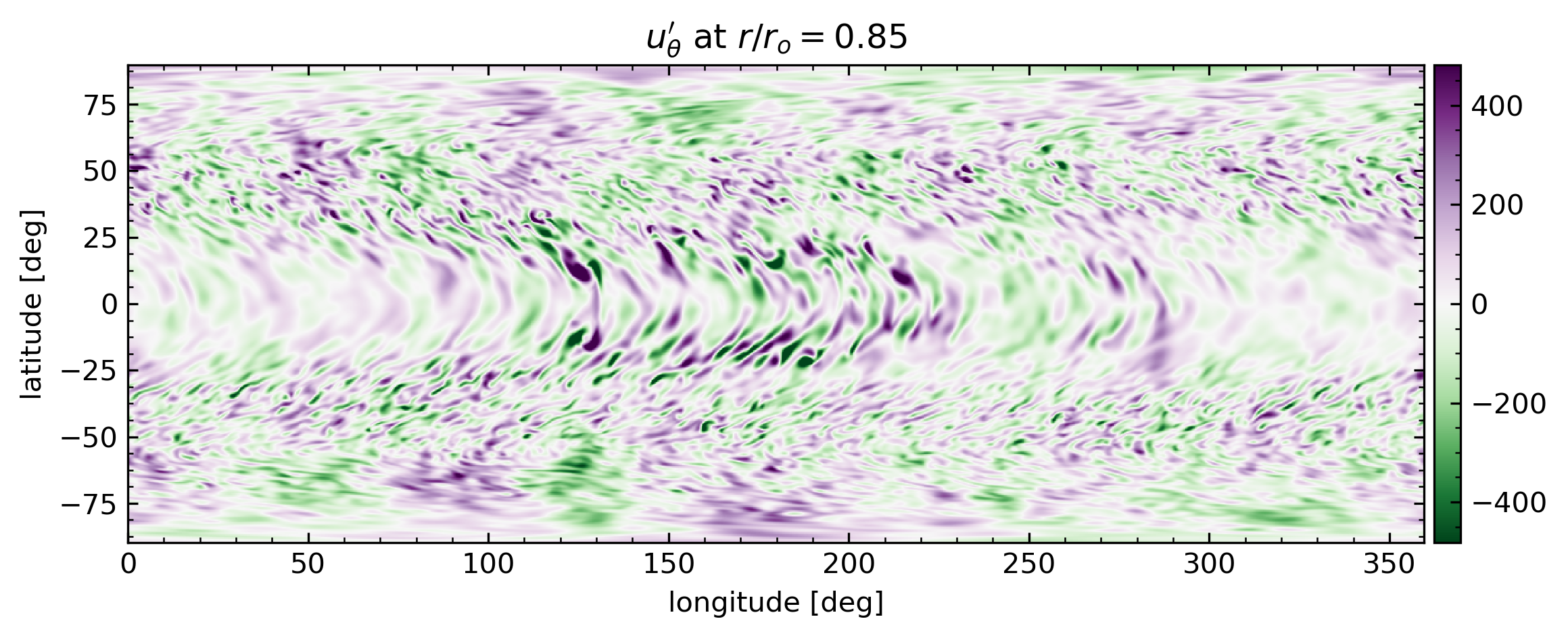}
    \caption{Snapshots of the statistically steady state at the shell center ($r/r_o=0.85$). The fluctuations around the meridional flow, $u_\theta'$, clearly show structures of $m\approx 5$ at higher latitudes.}
    \label{figAppHexagonlike}
\end{figure*}

These features also appear in snapshots of our simulation (see the region of latitudes between $\pm50$ and $\pm65$ degrees in both hemispheres in Fig.~\ref{figAppHexagonlike}). At these latitudes, structures in $m=3,\ldots,8$ are visible in both $u_\theta'$ (shown) and $u_\varphi'$ (not shown) that appear phase-shifted so as to form vortices.

\begin{figure}
    \centering
     \includegraphics[height=0.3\textwidth]{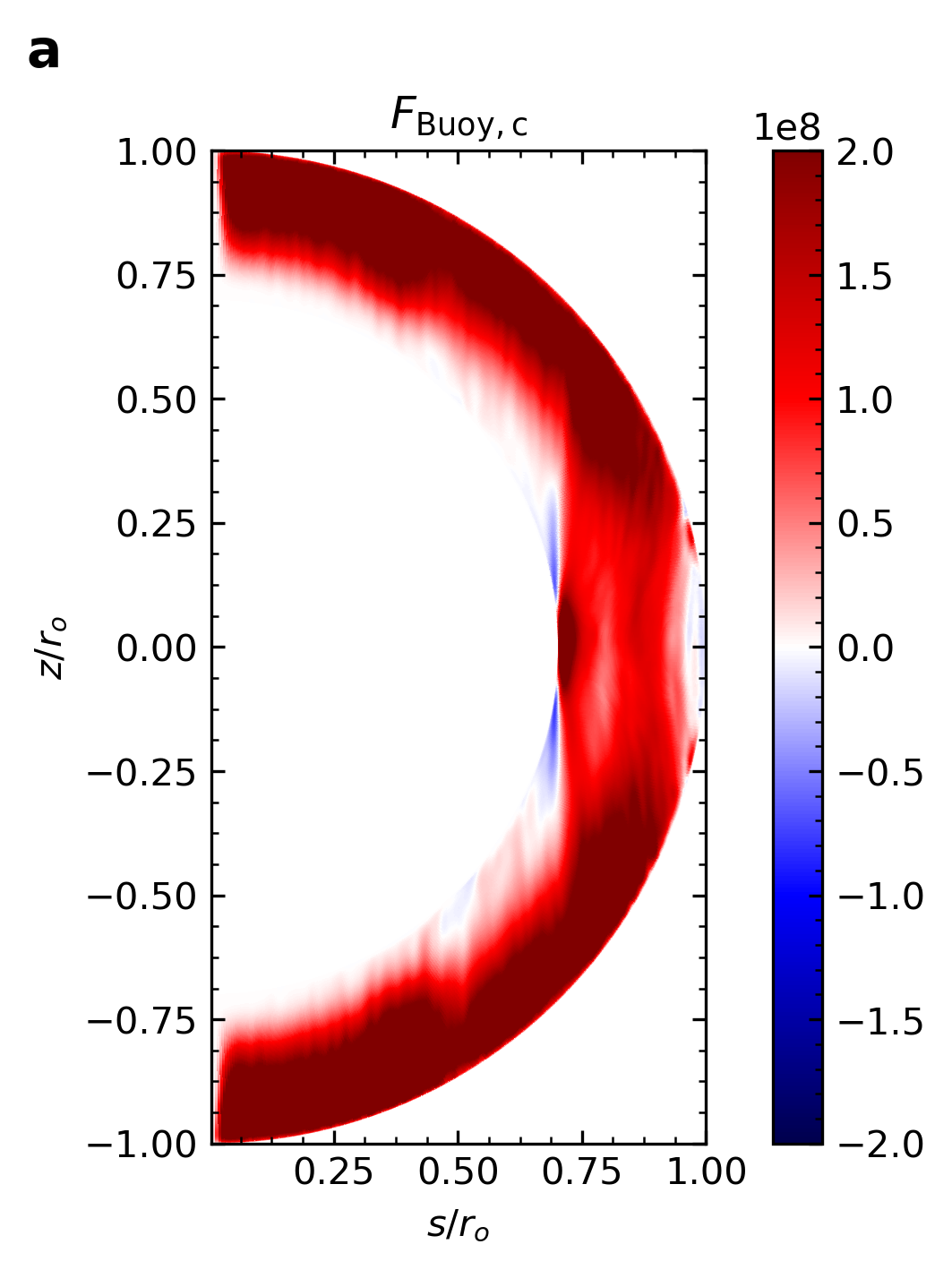}
     \includegraphics[height=0.3\textwidth]{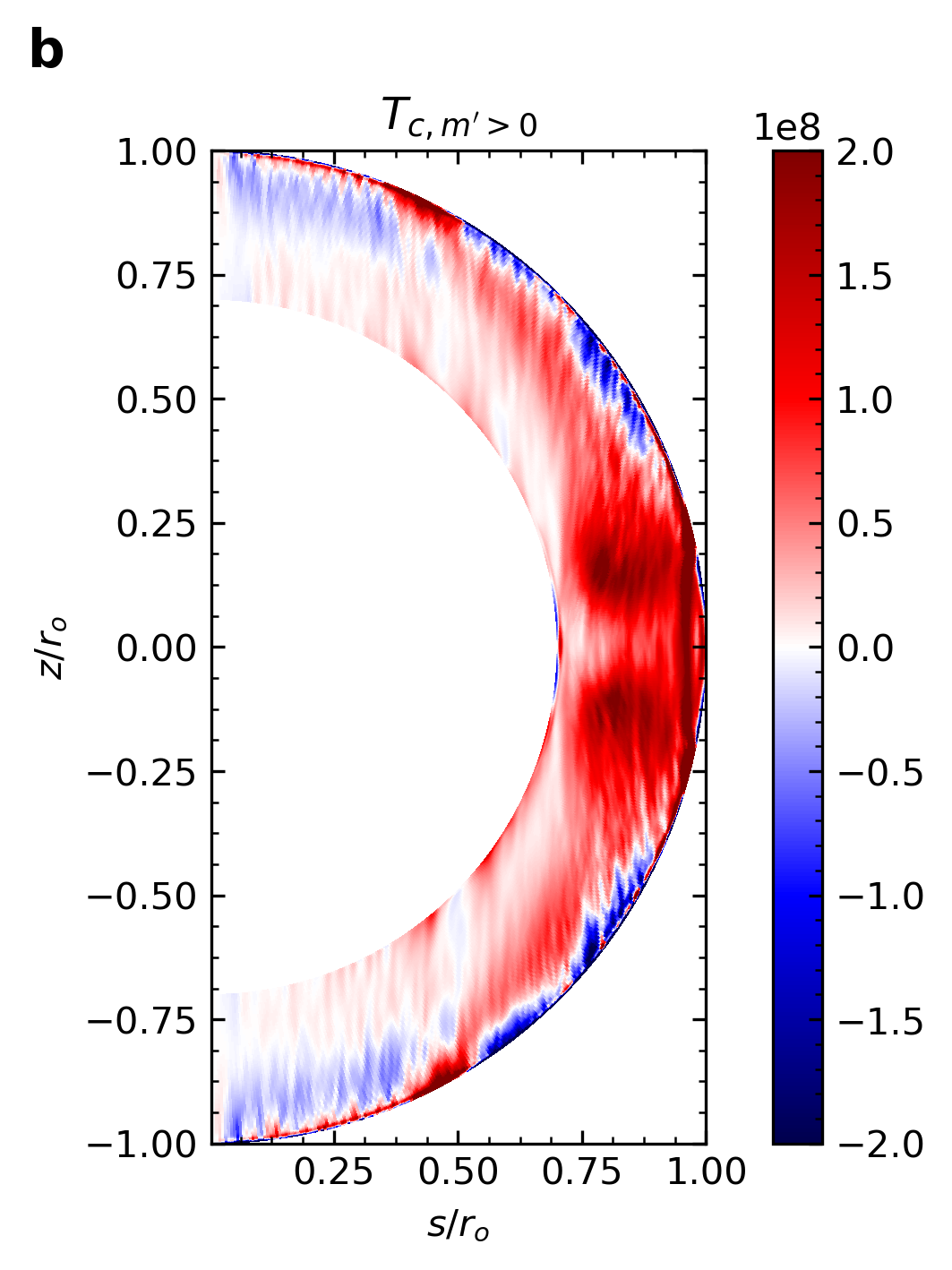}
    \caption{Spatial dependence of the driving of kinetic energy of large-scale eddies by buoyancy, $F_{\rm Buoy,c}$ (a), and kinetic energy transfer to the same eddies from other eddies, $T_{c,m'>0}$ (b), both in the statistically steady state. For both panels, we summed $T$ and $F_{\rm Buoy}$ over $m$ as given by label (c) in Fig.~\ref{figTmm} and summed $T$ over $m'>0$.}
    \label{figtMinusfBuoy}
\end{figure}

Over the entire shell, the transfer of kinetic energy from the jets to these large-scale vortices is smaller than the work done by buoyancy in our simulations (see a comparison between $F_{\rm Buoy}(m)$ and $T(m)$ at $3\leq m \leq 9$ in Fig.~\ref{figTPi}). In the region of the lower part of the shell at the retrograde jet, where the $m=3,\ldots,8$ wave numbers are strongest, however, the energy transfer due to $T(m,0)$ is larger than the driving by buoyancy. This is evident from a comparison of Figs.~\ref{figTrtheta}c and \ref{figtMinusfBuoy}a.

Figure~\ref{figtMinusfBuoy}b shows the spatial dependence of the transfer to the large-scale vortices from all other scales $m'>0$,
\begin{align}
T_{m'>0}(m)=\sum_{m'>0}T(m,m') \label{eqtmpgt0}.
\end{align}
We find that the driving of the large-scale vortices by other scales except the jet scale is weaker than the driving from the jets in the deeper regions, but stronger in the near-surface and equatorial regions.

We conclude that the deeper large-scale vortices are predominantly driven by a mechanism that transfers energy from the jets to these scales. The origin of this transfer of kinetic energy from the jets to the large-scale vortices is likely an instability of the jets. {We checked the zonal flow instability criterion from \citet[][which is the same as for barotropic instability]{Wicht2002b}, which however was not fulfilled and this instability can be ruled out.}

\subsubsection{Forward cascade to small scales}

Toward the classical dissipation scales ($m,m'>60$), we find {in agreement with the energy flux in Fig.~\ref{figTPi}} that kinetic energy is transferred to ever smaller scales (region (d) in Fig.~\ref{figTmm}). This transfer occurs in almost all of the spherical shell and in a stronger way toward the surface, where convective motions experience the strongest driving (see Figs.~\ref{figTrtheta}d and~\ref{figtMinusfBuoy}a).

The logarithmic axes in Fig.~\ref{figTmm} give the impression that the width of the region of forward transfer shrinks with $m$. However, this is not the case and the width of this region is indicative of the convective driving. The forward transfer is not strictly local but slightly nonlocal, meaning not only to directly neighboring scales but mostly to scales in a close neighborhood. Such a somewhat nonlocal transfer is expected for a driving that is broad in wavenumber space like our driving by buoyancy (see $F_{\rm Buoy}$ in Fig.~\ref{figTPi} and \citealp{Kuczaj2006}). For turbulent convection, similar results have been observed in {several Cartesian} studies \citep[e.g.,][]{Mishra2010,Moll2011,Verma2017,Alexakis2018} in both rapidly rotating and nonrotating systems \citep[e.g.,][]{Favier2014,Verma2017}. The non-locality of the transfers (i.e., the distance to the diagonal) can be explained by the necessity of the two wave numbers $m$ and $m'$ to have a coupling partner, for which only $m-m'$ and $m+m'$ are an option. 
A closer inspection of Fig.~\ref{figTmm} shows that the main coupling partners for $m$ are around $m'\approx m\pm 35$, which is the mean distance of region (d) from the diagonal. This can be explained by the main convective driving and main convective power being at the same scales $m-m'=m''\approx 35$ (see the poloidal spectrum in Fig.~\ref{figSpectra}b and $F_{\rm Buoy}$ in Fig.~\ref{figTPi}=. The forward transfer region observable in our simulation thus resembles a relatively local forward cascade \citep[e.g.,][]{Alexakis2018}.

{
\subsubsection{Local forward transfer}

Furthermore, there is a very narrow region of forward transfer of kinetic energy close to the diagonal, which at first sight looks like the sign of a forward cascade (region (e) in Fig.~\ref{figTmm}). Since this pattern starts at $m\sim10$ and since it strongly involves the interaction with the $m=1$ component, we suppose that this feature is related to the dynamics of the depth-averaged geostrophic component of the flow \citep[see][]{Rubio2014}. This is supported by panel (e) in Fig.~\ref{figTrtheta}, which shows that the interaction is due to a very large-scale component with a strong alignment in $z$. Similar results appear in \citet[][Fig. 7a,c]{Maltrud1993} in forced two-dimensional turbulence and {in} \citet[][bottom left panel in Fig.~5]{Rubio2014} in rapidly rotating Rayleigh-Bénard convection. These authors attributed similar features to the signature of a forward enstrophy transfer in the energy transfer function. In two-dimensional flows, the domain-integrated enstrophy $Z=\frac{1}{2} |\nabla \times \bu|^2$ is an invariant in addition to the kinetic energy and also shows forward cascading behavior \citep[e.g.,][]{Kraichnan1967,Alexakis2018}. The forward enstrophy cascade shows up in the kinetic energy transfer as a local forward transfer \citep[see also][]{Verma2005}, but without contributing to the kinetic energy flux (see discussion in \citealp{Maltrud1993} and \citealp{Kraichnan1967}). In agreement with this, we find that this region does not have a significant contribution to the scale-to-scale flux. Forward enstrophy transfer likely results from stretching of vortices by the background large-scale shear and is associated with the upscale flux of energy to the largest scales \citep[e.g., compare to ][]{Maltrud1993,Chen2006,Xiao2009}. In line with this, we find this transfer in our simulation to be strongest in the region of strongest shear in the zonal flow (see panel (e) in Fig.~\ref{figTrtheta}). These arguments support an attribution of this feature to forward enstrophy transfer. A separate analysis of the spectral transfer of enstrophy is necessary to firmly conclude this, which is left to future studies.

}

\section{Results for the spin-up phase}
\label{secResultsSpinup}

Once zonal flows have been established, the direct driving via Reynolds stresses dominates the feeding of energy into the jets. However, the mechanism that initializes the runaway growth could have a different origin. \citet[e.g.,][]{Busse2002} speculates that the boundary curvature establishes a small initial tilt because Rossby waves in the equatorial region tend to drift faster in prograde direction than at depth. This could explain the simple structure typically found outside the TC in simulations of convection in a deep rotating shell. In these simulations, a prograde equatorial jet dominates a flanking retrograde jet. However, for the multiple jets developing in many simulations inside the TC, a more complex explanation is required. Busse originally envisioned nested layers of convective columns, but this has never been confirmed in any numerical simulation.

The Rhines scale more or less correctly predicts the jet width inside but not outside the TC \citep{Gastine2014}. This could indicate that {the jet formation mechanisms are different inside and outside the TC. Inside the TC,} an initial cascade {could provide} the jet structure{ while outside TC, it is not present}. Once the jets are strong enough, however, the direct driving via Reynolds stress would dominate due to the dominating effect of the zonal flows on the dynamics {inside and outside the TC. A similar} behavior has for example been observed in forced two-dimensional turbulence on a sphere \citep{Nozawa1997,Nozawa1997b,Huang1998}.

In order to explore this possibility, we run a spin-up simulation that starts from an initial state with developed convective flows but without jets and evolves for about 0.21 viscous timescales. The initial condition is established somewhat artificially by forcing the axisymmetric contributions of flow, entropy and pressure to remain zero for a period of 0.15 viscous timescales. The exception are the spherical{ly} symmetric contribution to entropy and pressure, which describes the diffusive background state. The manipulated simulation was long enough to establish a new statistically steady state.

\begin{figure*}
    \centering
    \includegraphics[width=0.98\textwidth]{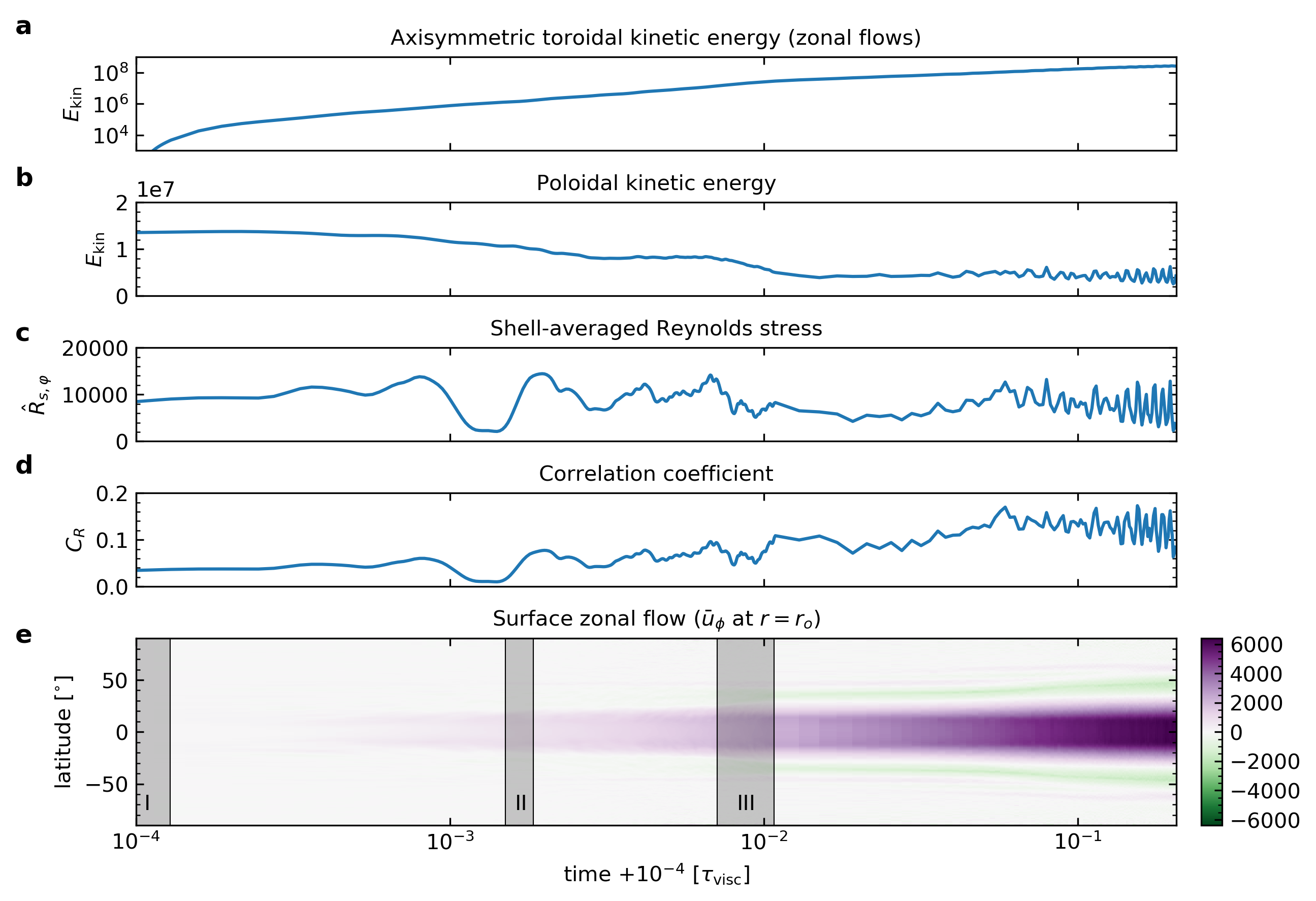}

    \caption{{Temporal development of the jets during spin-up. We show the kinetic energy of zonal flows (a) and poloidal flows (b, mostly convection), shell-averaged r.m.s. Reynolds stresses (c and d), and the surface jet profile (e).} See Eqs.~\eqref{eqReyAvg} and~\eqref{eqCReyAvg} for definitions of the spatial averages of the Reynolds stresses and correlation coefficients. This run was performed using a convective initial condition without jets. The duration of a convective turnover time is about $10^{-3} \tau_{\rm visc}$. We show results for the spatially resolved Reynolds stresses and the energy transfer to the jets during the intervals marked as gray bands in Figs.~\ref{figSpinupReynolds} and~\ref{figSpinupM0restart}.}
    \label{figButter}
\end{figure*}

Figure~\ref{figButter} gives an overview how the total zonal flow energy (a), the surface zonal flows, the poloidal energy (b), the volume-averaged Reynolds stress (c) and correlation $C_R$ of convective flows (d), and the surface zonal jets (e) develop during the spin up. The volume-averaged Reynolds stress \citep[][]{Christensen2002,Gastine2012},
\begin{align}
    \hat R_{s,\varphi} &= \frac{1}{V} \iint \tilde\rho\, \overline{u_s' u_\varphi'} \, r^2\sin\theta \, \id\theta \, \id r \label{eqReyAvg}, 
\end{align}
where $V$ is the shell volume, relies on the correlation of non-axisymmetric flow contributions in $s$ and the azimuthal direction,
\begin{align}
    C_{R} &= \frac{\iint \tilde\rho\,\overline{u_s' u_\phi'} \, r^2\sin\theta \, \id\theta \, \id r}{\sqrt{\iint \tilde\rho\,\overline{u_s' u_s'} \, r^2\sin\theta \, \id\theta \, \id r \,\iint \tilde\rho\,\overline{u_\varphi' u_\varphi'} \, r^2\sin\theta \, \id\theta \, \id r}} \label{eqCReyAvg}.
\end{align}
The manipulated initial state is characterized by particularly strong convective motions (b), which are suppressed during the spin-up by the growing zonal flows.

\begin{figure*}
    \centering
    \includegraphics[height=0.42\textwidth]{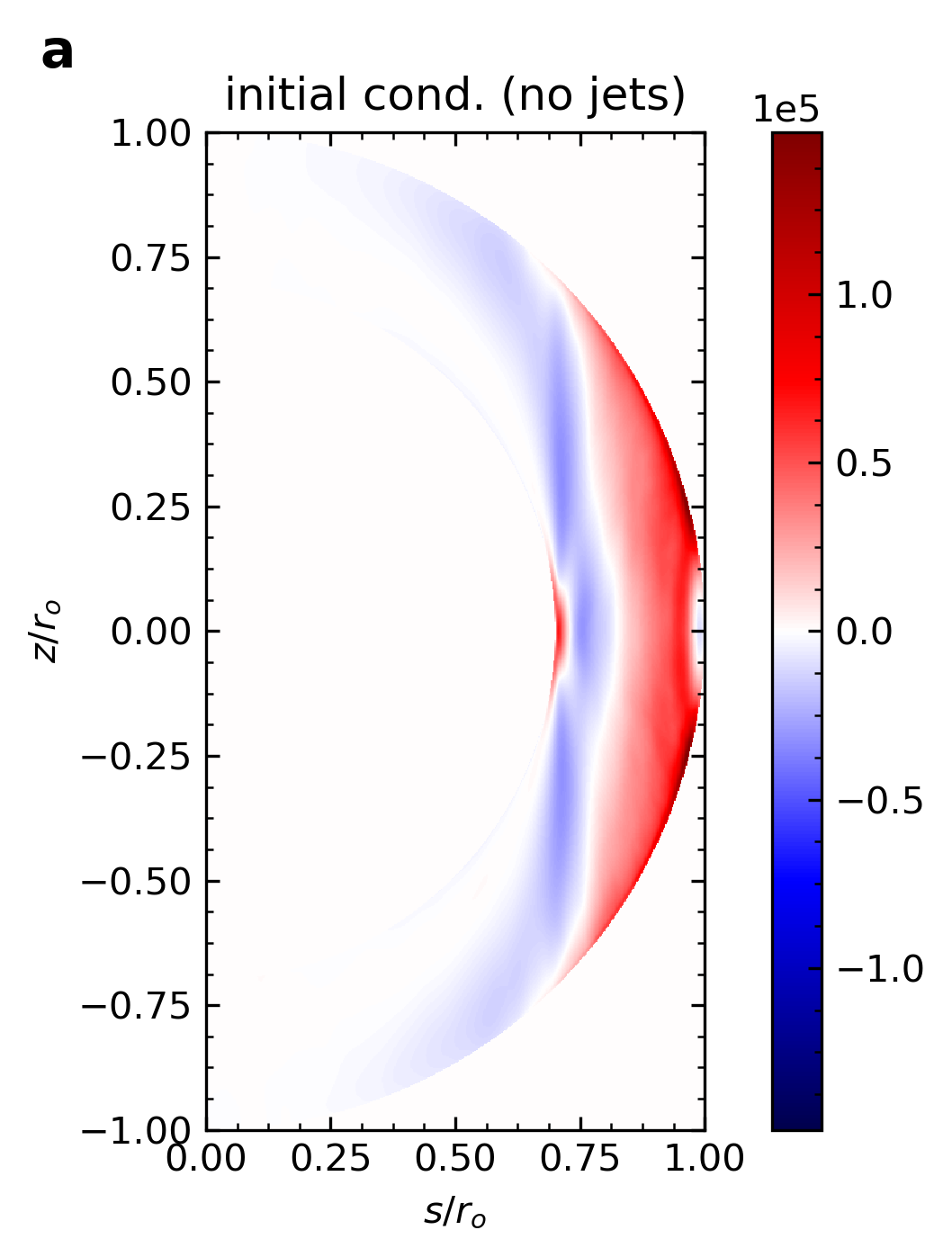}
    \includegraphics[height=0.42\textwidth]{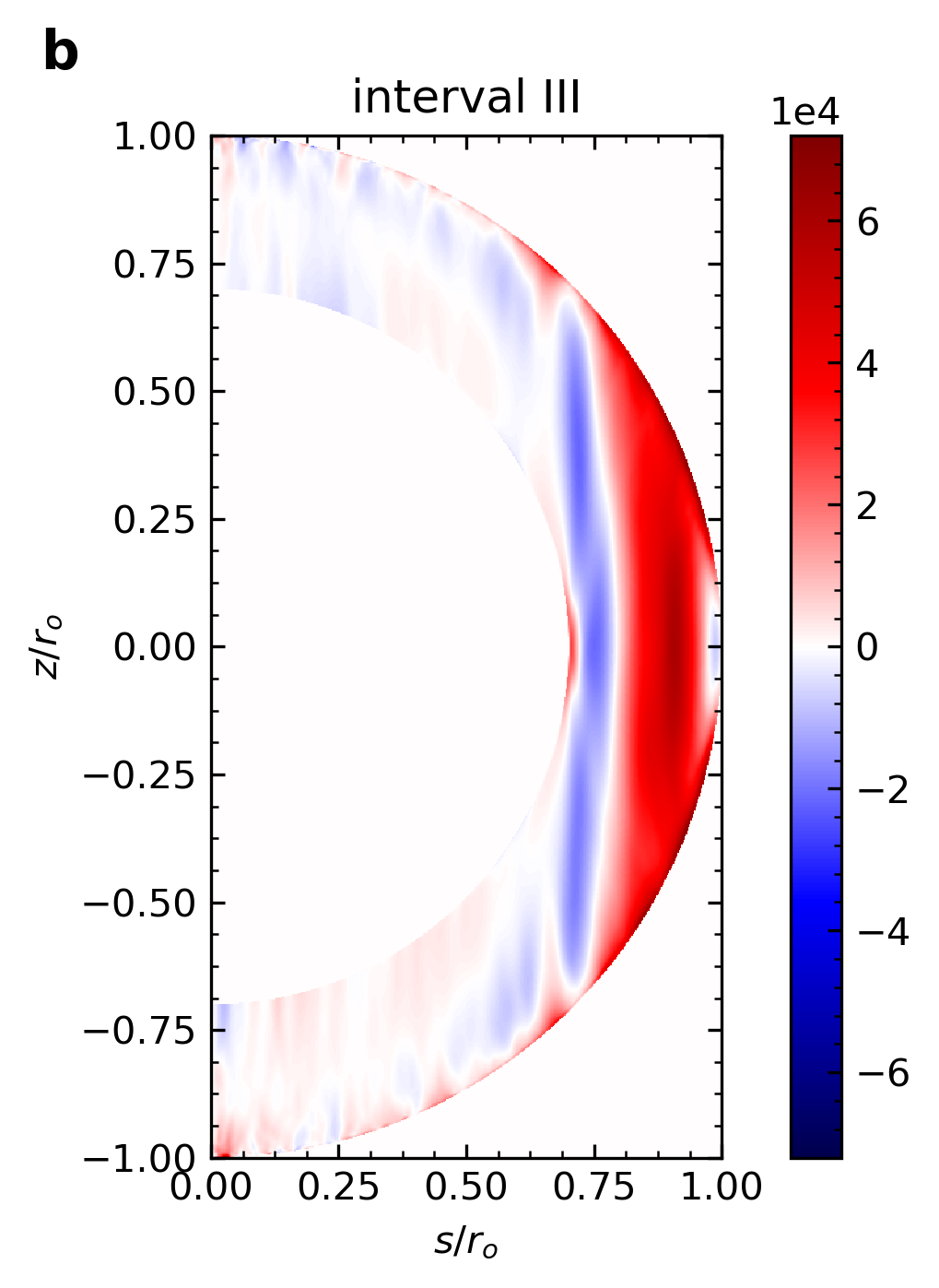}
    \includegraphics[height=0.42\textwidth]{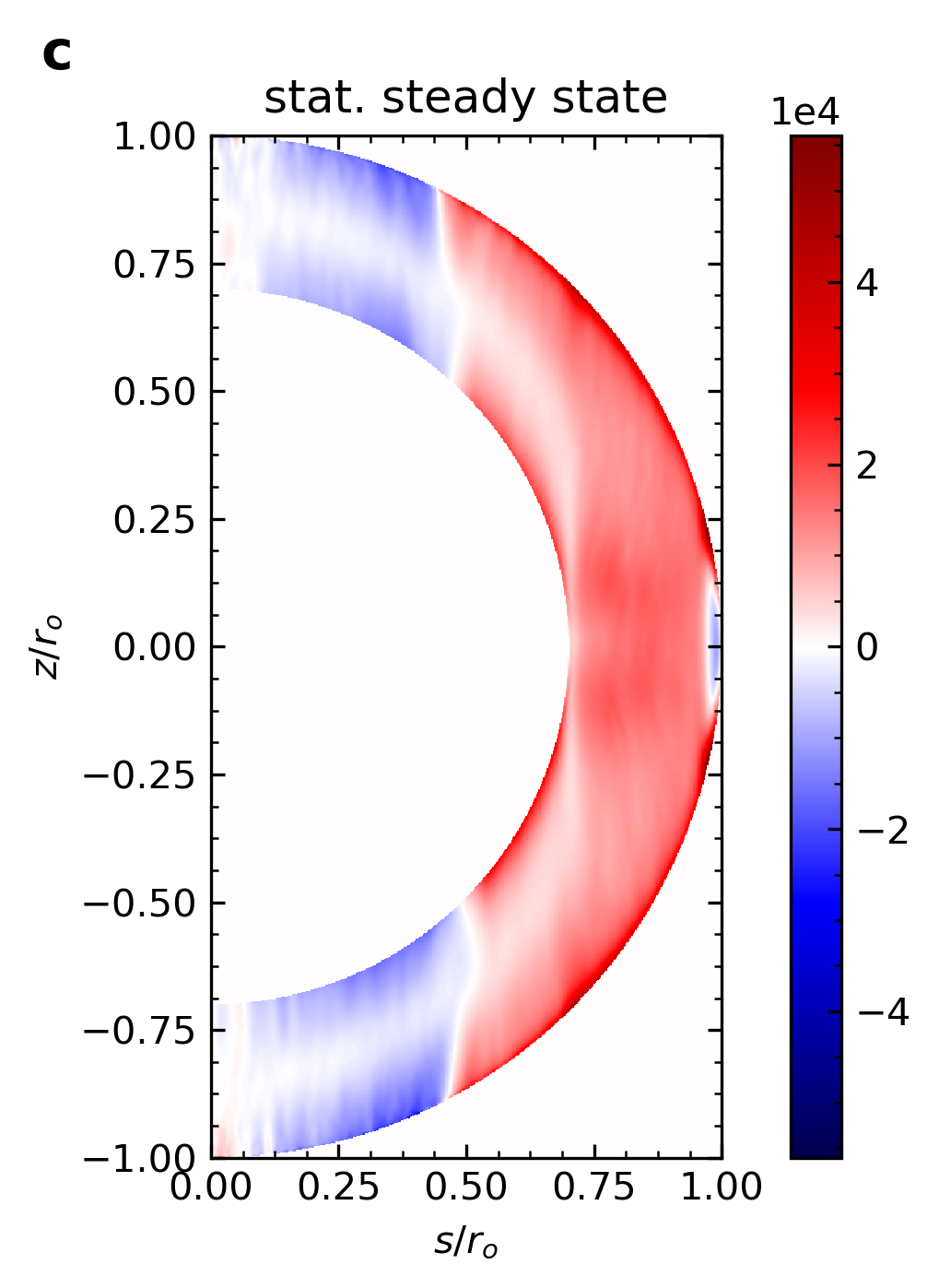}
    \caption{Reynolds stresses $R_{s,\varphi}$ at different times during the spin-up: at the initial condition without jets and uniform background rotation (a), after about 100 turnover times and in the growth phase of the jets (b, same as interval III in panel (e) in Fig.~\ref{figButter}) and in the statistically steady state with fully developed jets (c).}
    \label{figSpinupReynolds}
\end{figure*}

Even without zonal flows, the manipulated simulation shows a correlation $C_R$ and thus some Reynolds stress. The pattern of this Reynolds stress is illustrated in Fig.~\ref{figSpinupReynolds}a. Correlation and stress are mostly positive (red) in the region outside the TC. This is consistent with the theory outlined by \citet{Busse2002} that Rossby waves would travel faster in prograde direction closer to the equator and therefore cause a consistent tilt with a positive correlation $C_R$. Inside the TC, the Reynolds stress is much weaker and the correlation is mostly negative. The negative sign is also consistent with the Rossby wave theory, which would travel retrograde inside the TC and faster closer to the TC. We thus speculate that this Rossby wave theory largely explains the initial correlation when no zonal winds are present.

Figure~\ref{figButter} demonstrates that jets first appear outside the TC where the Reynolds stress is strongest. Over the spin-up process, the correlation increases (d) along with the zonal winds (a and e). The Reynolds stress amplitude (c), however, decreases, since the convective flow (b) becomes weaker.

Figure~\ref{figSpinupReynolds}b shows that the Reynolds stress distribution after about ten turnover times (or one percent of a viscous dissipation time) remains largely unchanged. However, at the end of the spin-up process (panel c), the pattern close to and inside the TC has clearly changed. The direction of the tilt of convective feature and thus of the correlation now changes roughly where the retrograde jets have their peak. The initial correlation and tilt structure is thus not the only factor that determines the end result. The simple Rossby wave picture may explain the initial correlation but fails to explain the evolution, at least inside the TC.

\begin{figure}
    \centering
    \includegraphics[width=\linewidth]{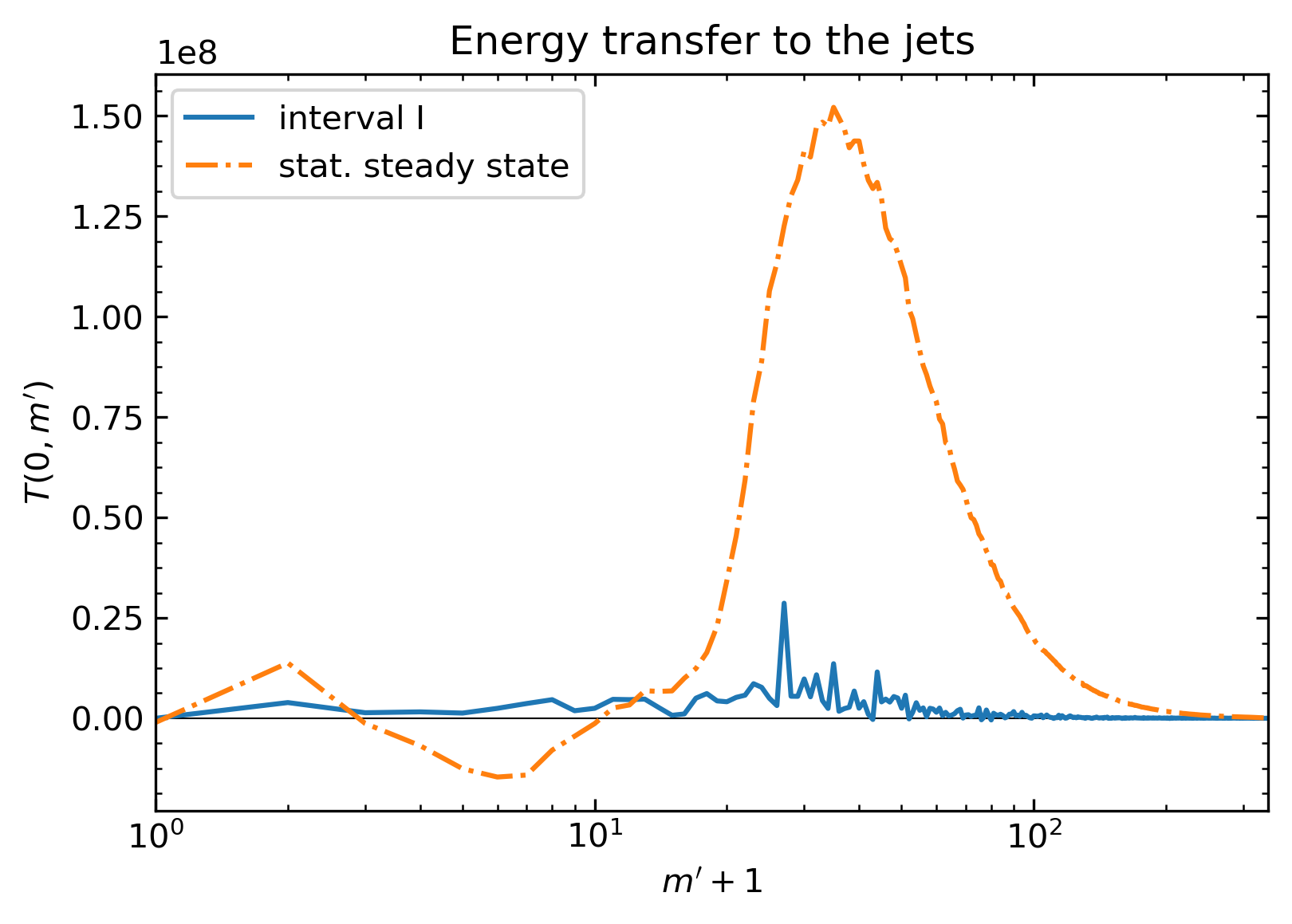}

    \caption{Kinetic energy transfer to the jets, $T(0,m')$, for the spin-up run. The transfer functions were computed for the first interval directly after the start of the spin-up (interval I). 
    The time window is shown as light gray band in Fig.~\ref{figButter}. For comparison, we also show results for the statistically steady state from Fig.~\ref{figTmm}.}
    \label{figSpinupM0restart}
\end{figure}

Figure~\ref{figSpinupM0restart} shows the energy transfer to the jet scale, $T(m=0,m')$, for the first time interval directly after the start of the spin-up. This interval has a short duration of less than a convective turnover time, which is at the order of $10^{-3}$ viscous timescales. During this period, the flows thus largely stem from the initial condition. We find that almost all azimuthal scales contribute to the initial growth of the jets, with a major contribution from the small convective scales (here around $m\approx 25$; slightly larger scales compared to the statistically steady state). This shows that from the start of the spin-up, the jets are not driven by an inverse cascade, but rather by direct feeding from convective structures. During the spin-up, the pattern of the transfer function evolves with the merging and growth of the jets to its steady-state version as the Reynolds stresses evolve (see the plot for the steady-state energy transfer in Fig.~\ref{figSpinupM0restart}). We also computed the transfer function $T(m,m')$ for $m,m'>0$ during spin-up (not shown). In the early phase of the spin-up, the results are too noisy to permit a conclusion because they have been averaged over a too short period in time (intervals I and II). At the later stages, when patterns appear in the transfer function, they are qualitatively rather similar to the steady state (Fig.~\ref{figTmm}). Some temporal evolution takes place; the transfer from the jets to large-scale vortices (label c in Fig.~\ref{figTmm}) appears for example only at the latest stage.

\section{Conclusions}
\label{secConlusions}

In this study we have determined the spectral transfer of kinetic energy in rapidly rotating spherical-shell convection for moderately turbulent {flows}. We ran the simulations until a statistically steady state was safely reached and studied the resulting statistical properties of the fluid motions to understand whether an inverse cascade may power jets and large-scale vortices.

We find rich dynamics that depend strongly on the azimuthal wave number, that is, on the azimuthal length scale. The dominant energy transfer is upscale and goes directly from the small convective scales into the jets. It is due to statistical correlations of the convective flows (Reynolds stresses). This transfer is very nonlocal in wavenumber space and thus not due to an inverse cascade, although it drives a relatively constant upscale scale-to-scale flux at large scales.

The energy transfer to the large-scale vortices is about an order of magnitude smaller than the transfer to the jets. 
Large-scale vortices receive energy through different mechanisms. The main driver is buoyancy, but there is also upscale transfer from small-scale convection and forward transfer from the jets. Outside the TC, the upscale transfer is again directly from the small convective scales and thus likely not due to an inverse cascade, although it does contribute to an upscale scale-to-scale flux. Inside the TC, the upscale transfer is more local, and we therefore cannot rule out an inverse cascade as a driver of large-scale vortices. The transfers inside the TC are, however, quite weak. The forward transfer from the jets {to large-scale vortices} mostly happens in deeper regions {and goes} to specific wave numbers around $m=5${. This transfer takes place predominantly in} a region at depth close to the jet center of the retrograde jet, where kinetic energy shows a peak {at the same wave numbers around $m=5$}. We therefore propose an instability of the jets as an alternative mechanism for the driving of large-scale vortices. 

We also find two qualitatively different regions of relatively local forward transfer. As expected, kinetic energy is transferred from the small convective scales to the classical dissipation scales in a forward cascade. 
In addition, we find a region of very local forward transfer that does not have a significant contribution to the scale-to-scale energy flux. This is likely the signature of local forward enstrophy transfer, a question that is left to future studies.

Analyzing the spin-up and growth of the jets more closely, we show that the growing jets suppress convection to some extent but result in an increased correlation in the convective flows, which is likely due to the growing shear. The increasing correlation compensates for the suppression of the convective flows and helps maintain strong Reynolds stresses, which drive the jets. 
The initial number of jets is continuously reduced by the merging of the jets. This continues until a balance with (turbulent) viscosity is reached.

We used a manipulated initial condition with developed convective flows but where the jet growth has been suppressed for a long simulation time. In this initial state, correlations of the small-scale convection remain due to the effect of uniform rotation and shell geometry. They drive the Reynolds stresses that kick off the jet growth. 
However, flow correlation and Reynolds stresses change over the spin-up process. While the initial correlation seems consistent with the Rossby wave picture of \citet[e.g,][]{Busse2002}, the evolution and final jet structure obviously require an additional ingredient. Whether a cascade still plays a role remains unclear, at least inside the TC. There is no clear support for a cascade from our transfer analysis. 

Previous analyses of simulations with multiple jets indicate that the Rhines scale explains the jet width inside but not outside the TC. We speculate that the strong correlation caused by the consistent tilt of convective  features outside the TC dictates the length scale. Consequently, a direct driving of the jets always dominates. Inside the TC, additional factors come into play, but the applicability of the Rhines scale does not necessarily imply the action of a cascade. The results from the spin-up may be used in future studies to improve rotating-turbulence models of stellar differential rotation and planetary jets, for example using the so-called $\Lambda$ effect \citep[e.g.,][]{Ruediger1989,Kitchatinov2013review,Barekat2021}.

We find that the classical picture of a step-by-step transfer of kinetic energy from one neighboring scale to another, as in a true inverse cascade, is not a driver of jet formation in our simulation. Instead, eddy-mean flow interactions are the main drivers of the jets, with eddy-eddy interactions playing only a minor role, which supports the results of quasi-linear simulations  \citep{Tobias2011,Srinivasan2012,Marston2016}. Our results are consistent with the dominance of eddy-mean flow interactions over eddy-eddy interactions in observations \citep[][]{Young2017}. This result may be underpinned by a comment by \citet[][p. 251]{Frisch1995TurbulenceBook}, who stated that an inverse cascade is typically associated with unbounded domains, as the generation of coherent vortical structures at the finite box size seems to ruin the classical picture of the inverse cascade. Accordingly, simulations of inverse cascades in a box need to suppress the growth of the box-scale mode via a strong damping of the largest scales \citep[e.g.,][]{Chen2006,Xiao2009} to actually obtain an inverse cascade. This is also in agreement with simulations of rapidly rotating Rayleigh-Bénard convection \citep[e.g.,][]{Rubio2014,Favier2014}, where energy accumulates at two box-filling vortices that are in the statistically steady state and directly driven by upscale transfers from the injection scale. These authors seem to associate {their} results with an inverse cascade despite {their finding of} nonlocal upscale transfers to the largest scale.

Here, we analyze{d} the transfer between azimuthal wave numbers, $m$, in order to be able to infer, in particular, the transfer to the jets and the dependence on convective scale as well as on latitude. In future studies, we may also analyze the spectral transfer in harmonic degree, $\ell$, as considered by \citet[][]{Young2017}. This may provide answers to the questions of how the jets merge in the simulations and what sets the resulting number of jets and their width \citep[e.g.,][]{Rhines1975,Gastine2014}. {Additionally, such a study may help explain the scaling of the {kinetic energy spectrum of the jets}, for which our results are similar to the $\ell^{-5}$ scaling in zonostrophic turbulence {(see, e.g., \citealp{Rhines1975,Sukoriansky2007,Galperin2019}). Broadly speaking, our simulation may well be compared to the zonostrophic regime, because of the $\ell^{-5}$ scaling and because the relevant spectral range is sufficiently large. On the other hand, we also find significant differences compared to the zonostrophic regime, most importantly the absence of an isotropic inverse cascade. This might be because the $\beta$ effect is already a relevant effect at the forcing scale. In future studies, it would therefore be interesting to study the spectral properties for simulations that span a larger range of parameters, including varying degrees of stratification that would lead to smaller forcing scales. Finally, we note that our simulations do not include a large-scale drag, unlike simulations of the zonostrophic regime \citep[see, e.g.,][]{Sukoriansky2007,Galperin2019}; instead, the viscous force is due to viscous friction in the bulk of the domain. It is therefore possible that the $\ell^{-5}$ scaling applies to a wider class of flows than previously thought.}

Our simulation of a rather thick spherical shell produces a main equatorial jet but only two flanking jets in each hemisphere. It would be interesting to analyze more Jupiter-like simulations in a thinner shell that yield a richer jet structure in the future. This would also help clarify the possible differences inside and outside the TC and help us understand what determines the jet width.

The shear exerted by fully developed jets always leads to a significant tilt of convective features. It is therefore no surprise that the direct driving of the jets dominates. This will very likely also be the case for a more complex jet system. 

Finally, an analysis of observational data would be very interesting, such as from the Cassini or Juno missions \citep[e.g.,][]{Galperin2014,Young2017,Siegelman2022}. A comparison between observations, deep spherical shell simulations, and shallow general circulation models \citep[e.g.,][]{Schneider2009,Cabanes2020} may help resolve the question of whether the jet driving is deep or shallow.

\begin{acknowledgements}

{The authors thank the referee, Peter Read, for very helpful and insightful comments which greatly improved the manuscript.} VB thanks P. Read, R. Yadav, Jin-Han Xie, K. Julien, M. Rheinhardt, A. Barekat, M. Käpylä, and J. Warnecke for discussions. VB thanks R. Cameron for pointing him to the topic of kinetic energy transfer. This work was supported by the Deutsche Forschungsgemeinschaft (DFG) within the Priority Program SPP~1992 ``The Diversity of Exoplanets". 
This work used NumPy \citep{Oliphant2006,vanderWalt2011}, matplotlib \citep{Hunter2007}, SciPy \citep{Virtanen2020}, and SHTns \citep[][]{Schaeffer2013}.

\end{acknowledgements}




\end{document}